\documentclass[iop,revtex4.1]{emulateapj}

\usepackage{natbib}
\usepackage{amsmath}
\usepackage{color}
\usepackage{url}
\usepackage{ulem}
\usepackage{multirow}
\usepackage{amsmath}
\usepackage{wasysym}
\usepackage{amssymb}
\usepackage{pifont}
\newcommand{\xmark}{\ding{53}}%

\bibpunct{(}{)}{;}{a}{}{,}

\def\apjl{ApJ}%
\def\apjs{ApJS}%
\def\ao{Appl.~Opt.}%
\def\aap{A\&A}%

\shorttitle{ {\it NuSTAR} unveils a low-luminosity heavily obscured AGN in the LIRG NGC 6286}
\shortauthors{Ricci et al.}

\begin{document}


\title{{\it NuSTAR} unveils a heavily obscured low-luminosity Active Galactic Nucleus
\newline
 in the Luminous Infrared Galaxy NGC 6286

}


%

\author{C. Ricci\altaffilmark{1,2,*}, F. E. Bauer\altaffilmark{1,2,3,4}, E. Treister\altaffilmark{5,2}, C. Romero-Ca$\tilde{\rm \textsc n}$izales\altaffilmark{1,3}, P. Arevalo\altaffilmark{6},  K. Iwasawa\altaffilmark{7}, G. C. Privon\altaffilmark{5}, D. B. Sanders\altaffilmark{8}, K. Schawinski\altaffilmark{9}, D. Stern\altaffilmark{10}, M. Imanishi\altaffilmark{11,12,13}
}

\altaffiltext{1}{Instituto de Astrof\'{\i}sica, Facultad de F\'{i}sica, Pontificia Universidad Cat\'{o}lica de Chile, Casilla 306, Santiago 22, Chile} 
\altaffiltext{2}{EMBIGGEN Anillo}
\altaffiltext{3}{Millennium Institute of Astrophysics, Chile} 
\altaffiltext{4}{Space Science Institute, 4750 Walnut Street, Suite 205, Boulder, Colorado 80301, USA}
\altaffiltext{5}{Universidad de Concepci\'on, Departamento de Astronom\'ia, Casilla 160-C, Concepci\'on, Chile}
\altaffiltext{6}{Instituto de F\'isica y Astronom\'ia, Facultad de Ciencias, Universidad de Valpara\'iso, Gran Bretana N¼ 1111, Playa Ancha, Valpara\'iso, Chile.}
\altaffiltext{7}{ICREA and Institut de Ci\`encies del Cosmos, Universitat de Barcelona, IEEC-UB, Mart\'i i Franqu\`es, 1, 08028 Barcelona, Spain}
\altaffiltext{8}{Institute for Astronomy, 2680 Woodlawn Drive, University of Hawaii, Honolulu, HI 96822}
\altaffiltext{9}{Institute for Astronomy, Department of Physics, ETH Zurich, Wolfgang-Pauli-Strasse 27, CH-8093 Zurich, Switzerland}
\altaffiltext{10}{Jet Propulsion Laboratory, California Institute of Technology, Pasadena, CA 91109, USA}
\altaffiltext{11}{Subaru Telescope, 650 North A'ohoku Place, Hilo, Hawaii, 96720, U.S.A.} 
\altaffiltext{12}{National Astronomical Observatory of Japan, 2-21-1 Osawa, Mitaka, Tokyo 181-8588, Japan}
\altaffiltext{13}{Department of Astronomical Science, The Graduate University for Advanced Studies (SOKENDAI),  Mitaka, Tokyo 181-8588, Japan}

\altaffiltext{*}{cricci@astro.puc.cl}

\begin{abstract}
We report the detection of a heavily obscured Active Galactic Nucleus (AGN) in the luminous infrared galaxy (LIRG) NGC 6286, identified in a 17.5\,ks {\it NuSTAR} observation. The source is in an early merging stage, and was targeted as part of our ongoing {\it NuSTAR} campaign observing local luminous and ultra-luminous infrared galaxies in different merger stages. NGC\,6286 is clearly detected above 10\,keV and, by including the quasi-simultaneous {\it Swift}/XRT and archival {\it XMM-Newton} and {\it Chandra} data, we find that the source is heavily obscured [$N_{\rm\,H}\simeq (0.95-1.32)\times 10^{24}\rm\,cm^{-2}$], with a column density consistent with being Compton-thick [CT, $\log (N_{\rm\,H}/\rm cm^{-2})\geq 24$]. The AGN in NGC 6286 has a low absorption-corrected luminosity ($L_{2-10\rm\,keV}\sim 3-20\times 10^{41}\rm\,erg\,s^{-1}$) and contributes $\lesssim$1\% to the energetics of the system. Because of its low-luminosity, previous observations carried out in the soft X-ray band ($<10$\,keV) and in the infrared did not notice the presence of a buried AGN. NGC\,6286 has multi-wavelength characteristics typical of objects with the same infrared luminosity and in the same merger stage, which might imply that there is a significant population of obscured low-luminosity AGN in LIRGs that can only be detected by sensitive hard X-ray observations.

\end{abstract}


\section{Introduction}

Luminous [$L_{\rm\,IR}(8-1000\,\mu\rm\,m)=10^{11}-10^{12}$ $L_{\odot}$] and ultra-luminous ($L_{\rm\,IR}\geq 10^{12}L_{\odot}$) infrared galaxies (LIRGs and ULIRGs, respectively) were first discovered in the late sixties \citep{Low:1968cr,Kleinmann:1970nx}. With the advent of the {\it Infrared Astronomical Satellite} ({\it IRAS}, see \citealp{Sanders:1996uq} for a review), which discovered a large number of U/LIRGs, the cosmological importance of these objects became evident. Although they are relatively rare at low redshift, their luminosity function is very steep \citep{Le-Floch:2005zr}, and they are the major contributor to the IR energy density at $z\simeq 1-2$ (e.g., \citealp{Caputi:2007kl}, \citealp{Goto:2010qa}).

The discovery that most, if not all, U/LIRGs are triggered by galaxy mergers led to the development of an evolutionary scenario \citep{Sanders:1988kl} in which two gas rich disk galaxies collide, triggering an intense phase of star formation in which they are observed as U/LIRGs. This is then followed by a blowout phase, during which most of the material enshrouding the supermassive black hole (SMBH) is blown away and the system is observed as a luminous red quasar (e.g., \citealp{Glikman:2015ys} and references therein). When most of the dust is removed, the system is eventually observed as a blue quasar. This model is consistent with the observed increase of the fraction of obscured sources with redshift up to $z\simeq 3$  \citep{Treister:2010fk,Ueda:2014uq}. The {\it Wide-field Infrared Survey Explorer} satellite ({\it WISE}) has recently found evidence of a new population of very luminous IR sources ($L_{\rm\,IR} > 10^{13} L_{\odot}$), dubbed Hot Dust Obscured Galaxies (Hot DOGs, \citealp{Wu:2012bh}), which might represent a short evolutionary phase in the evolution of galaxies, and be related to mergers (e.g., \citealp{Eisenhardt:2012ve,Stern:2014kx,Assef:2015zr}). Numerical simulations (e.g., \citealp{Springel:2005hc}) have also shown that tidal interactions can drive an inflow of material that triggers and feeds both accretion onto the SMBH and star formation. Therefore, mergers might play an important role in fuelling the SMBH, as it has been suggested by the discovery that the fraction of active galactic nuclei (AGN) in mergers increases with the AGN luminosity \citep{Treister:2012ys,Schawinski:2012zr}, spanning from $< 1\%$ at a 2--10\,keV luminosity of $L_{2-10}\sim 10^{41}\rm\,erg\,s^{-1}$ to 70-80\% in the most luminous quasars with $L_{2-10}\sim 10^{46}\rm\,erg\,s^{-1}$.

\begin{figure}
\includegraphics[width=0.35\textwidth,angle=270]{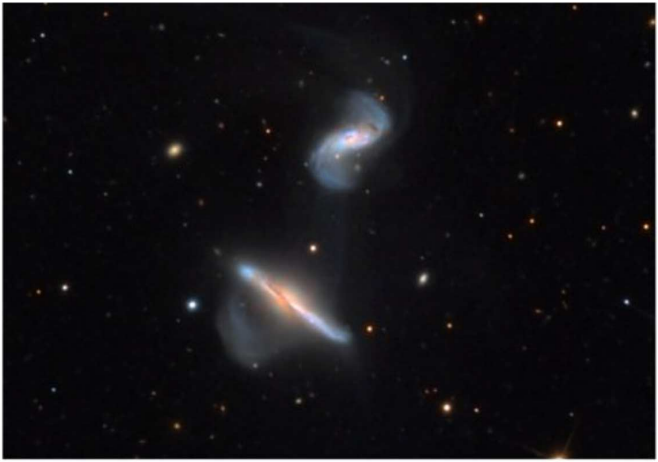}
\caption{ Optical image of the interacting pair NGC\,6286 (bottom)/NGC\,6285 (top) obtained with the Schulman 32-inch Telescope of the Mount Lemmon SkyCenter. Image courtesy of Adam Block (Mount Lemmon SkyCenter/University of Arizona).}
\label{fig:opticalimage}
\end{figure}

\begin{figure*}[t!]
\centering
\begin{minipage}[!b]{.48\textwidth}
\centering
\fbox{\includegraphics[width=8.5cm]{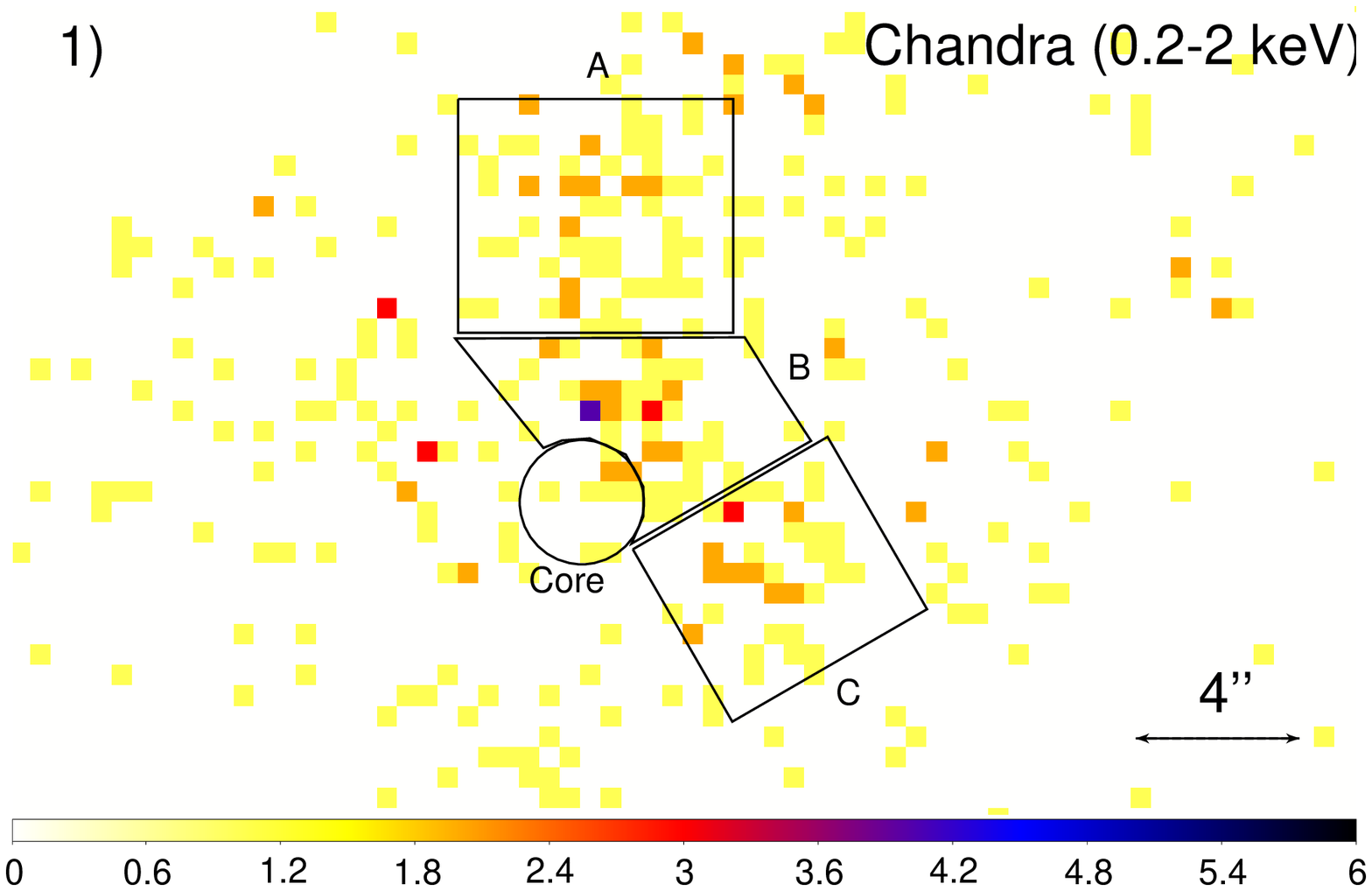}}\end{minipage}
\begin{minipage}[!b]{.48\textwidth}
\centering
\fbox{\includegraphics[width=8.5cm]{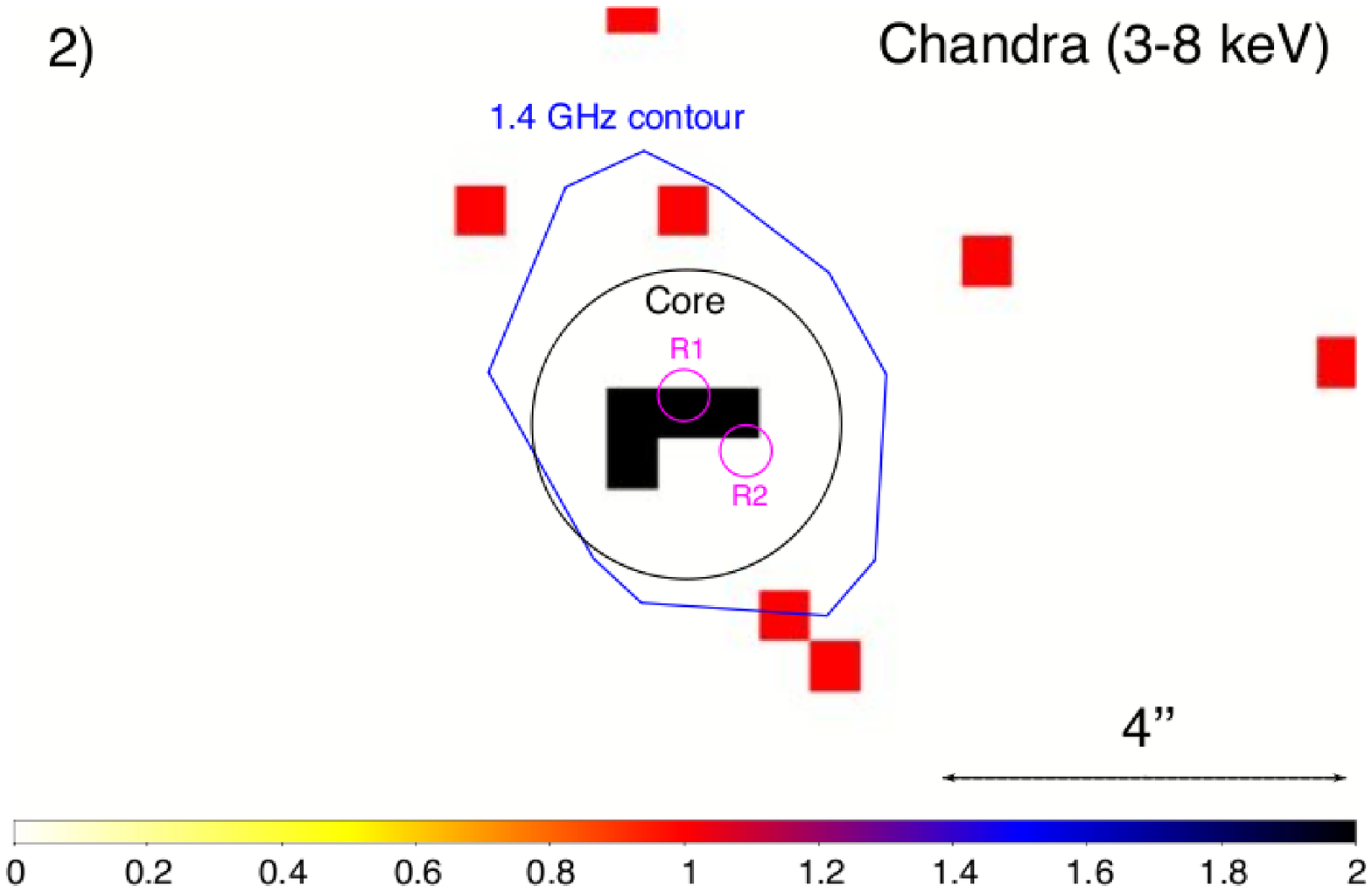}}\end{minipage}
\begin{minipage}[!b]{.48\textwidth}
\centering
\fbox{\includegraphics[width=8.5cm]{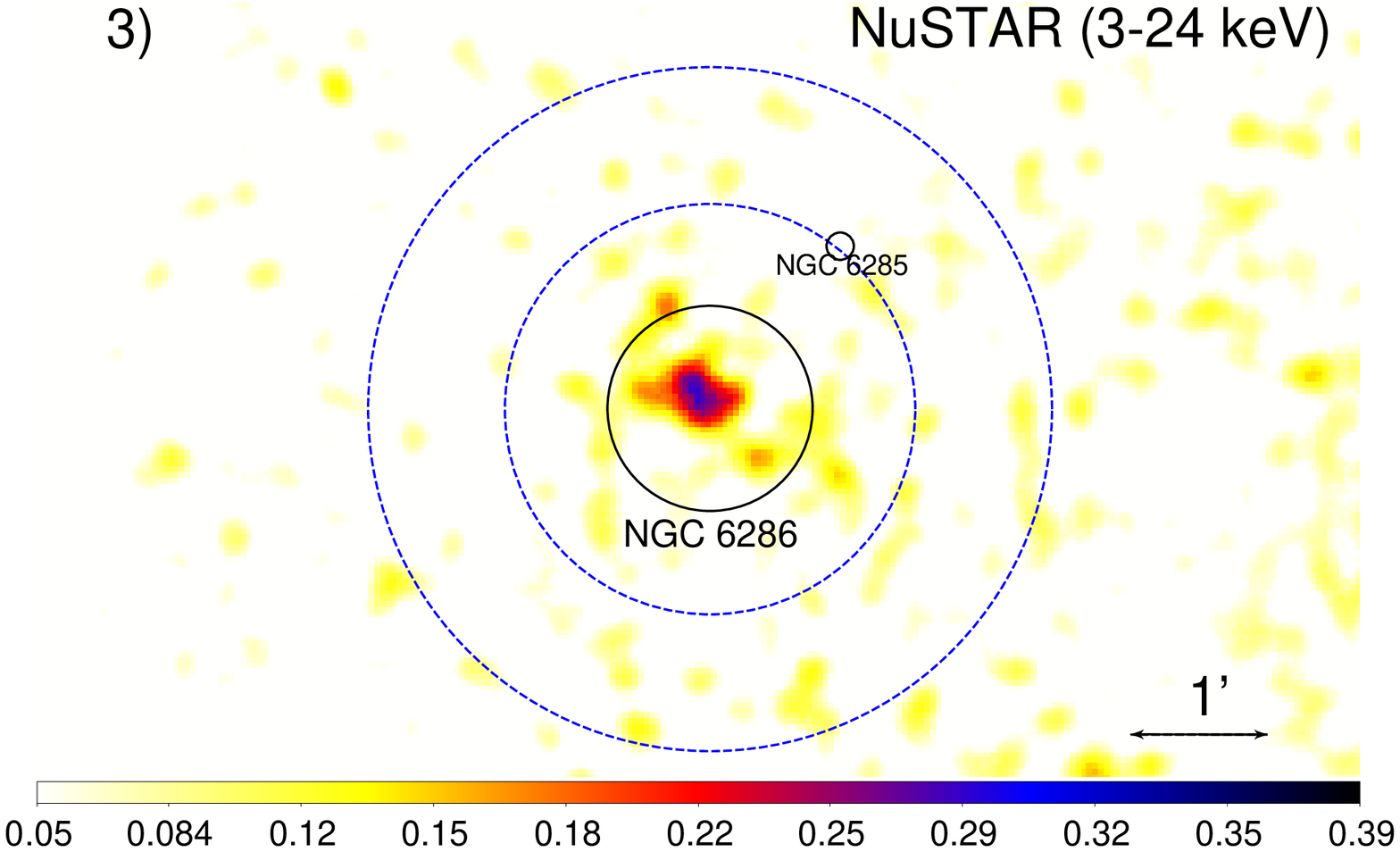}}\end{minipage}
\begin{minipage}[!b]{.48\textwidth}
\centering
\fbox{\includegraphics[width=8.5cm]{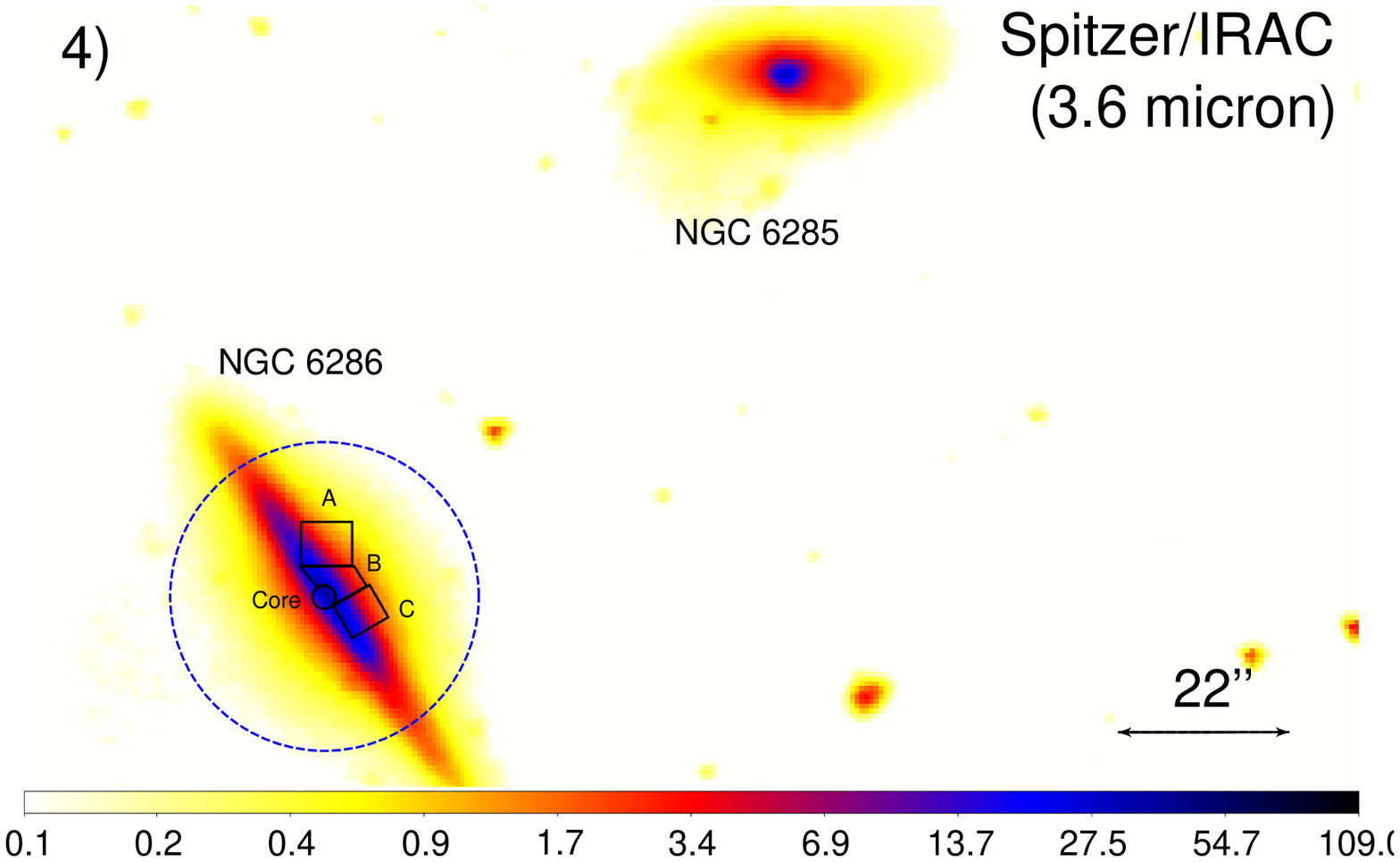}}\end{minipage}
 \begin{minipage}[t]{\textwidth}
  \caption{{\it Chandra} ACIS-S ({\it panel one}, 0.2--2\,keV; {\it panel two}, 3--8\,keV), {\it NuSTAR} FPMA ({\it panel three}, 3--24\,keV) and {\it Spitzer} IRAC ({\it panel four}, 3.6\,$\mu$m) images of NGC\,6286. The {\it Chandra} 0.2--2\,keV image shows a clear extended structure of $\sim 12$\,arcsec size ($\sim4.4$\,kpc). The four regions shown in panels one and four represent the 3--8\,keV core, the north, central, and south regions discussed in $\S$\,\ref{sect:specAnalysis_extended_nuclear}. The 1.4\,GHz VLA FIRST radio contour is illustrated in panel two, together with the two radio sources (R1 and R2) detected by our analysis of EVN observations at 5\,GHz (see $\S$\,\ref{Sect:radio}), which show that the radio emission coincides with the hard X-ray core, suggesting the presence of a buried AGN. The black circle and the blue dashed annulus in panel three correspond to the {\it NuSTAR} source and background extraction regions, respectively. The image in panel four was smoothed with a Gaussian kernel of radius 5 pixels. The blue dashed circle in panel four represents the source region used for {\it XMM-Newton} EPIC/PN.
    }
\label{fig:images}
 \end{minipage}
\end{figure*}

The contribution of AGN to the overall luminosity of U/LIRGs has been shown to increase with the IR luminosity of the system (e.g., \citealp{Veilleux:1995if,Veilleux:1999kc,Imanishi:2009ff,Imanishi:2010pi,Imanishi:2010uq,Nardini:2010dz,Alonso-Herrero:2012ud,Ichikawa:2014kx}).
Due to the great opacity of the nuclear region, a clear identification of AGN in U/LIRGs is often complicated. Mid-IR (MIR) properties have been used to estimate the relative contribution of accretion onto the SMBH and star formation to the bolometric luminosity. This has been done by exploiting 5--8\,$\mu$m spectroscopy (e.g., \citealp{Nardini:2010dz}), and the characteristics of several features in the $L$ (3--4\,$\mu$m) and $M$ (4--5\,$\mu$m) bands \citep{Imanishi:2000fu,Risaliti:2006kx,Sani:2008qa,Risaliti:2010kl}: the 3.3\,$\mu$m polycyclic aromatic hydrocarbon (PAH) emission feature, the bare carbonaceous 3.4\,$\mu$m absorption feature, and the slope of the continuum. The 6.2$\mu$m (e.g., \citealp{Stierwalt:2013ff,Stierwalt:2014lh}) and 7.7$\mu$m PAH features (e.g., \citealp{Veilleux:2009qo}), the presence of high-excitation MIR lines (e.g., [Ne\,V]\,$14.32\mu$m, \citealp{Veilleux:2009qo}), or the radio properties (e.g., \citealp{Parra:2010hc}, \citealp{Romero-Canizales:2012uq}, \citealp{Vardoulaki:2015mi}) have also been used to infer the presence of a buried AGN.

X-ray observations are a very powerful tool to detect accreting SMBHs and to disentangle the contributions of star formation and AGN emission to the total luminosity of U/LIRGs. Studies performed so far using {\it XMM-Newton} (e.g., \citealp{Franceschini:2003uq,Imanishi:2003vn,Pereira-Santaella:2011fu}) and {\it Chandra} (e.g., \citealp{Ptak:2003fv,Teng:2005bs,Iwasawa:2011fk}) have characterized the properties of a significant number of these systems. However, a significant fraction of U/LIRGs might be heavily obscured (e.g., \citealp{Treister:2010zr,Bauer:2010fk}), and X-rays at energies $\lesssim10$\,keV are strongly attenuated in Compton-thick (CT, $N_{\rm\,H}\geq 10^{24}\rm\,cm^{-2}$) AGN. Observations carried out in the hard X-ray band ($\geq 10\rm\,keV$) are less affected by absorption, and can be used to probe nuclear X-ray emission even in highly obscured systems (e.g., \citealp{Balokovic:2014dq}, \citealp{Gandhi:2014bh}, \citealp{Arevalo:2014nx}, \citealp{Bauer:2015ve}, \citealp{Koss:2015bh}, \citealp{Lansbury:2015dq}, \citealp{Annuar:2015qf}, \citealp{Puccetti:2015qf}, \citealp{Ricci:2015fk}). Previous hard X-ray observations of U/LIRGs have been carried out with {\it BeppoSAX} (e.g., \citealp{Vignati:1999ys}), {\it Suzaku} (e.g., \citealp{Teng:2009ij}) and {\it Swift}/BAT \citep{Koss:2013pi}.

The recent launch of the {\it Nuclear Spectroscopic Telescope Array} ({\it NuSTAR}, \citealp{Harrison:2013uq}), the first focussing telescope in orbit operating at $E\geq 10$\,keV, has opened a new window in the study of U/LIRGs thanks to its unprecedented characteristics. The first studies of the hard X-ray emission of local ULIRGs carried out with {\it NuSTAR} have recently been reported by \cite{Teng:2015vn} and \cite{Ptak:2015nx}, who show the importance of sensitive hard X-ray spectra to well constrain the line-of-sight column density.

 We report here on the first results of a series of {\it NuSTAR} observations awarded to our group during AO-1 as a part of a campaign aimed at observing ten local LIRGs in different merger stages (PI: F. E. Bauer). The sources were selected from the Great Observatories All-sky LIRG Survey (GOALS\footnote{http://goals.ipac.caltech.edu}, \citealp{Armus:2009fk}). GOALS is a local ($z<0.088$) sample which contains 181 LIRGs and 21 ULIRGs selected from the {\it IRAS} Revised Bright Galaxy Sample \citep{Sanders:2003fk}.

This paper reports the detection of a heavily obscured AGN in NGC\,6286 (also referred to as NGC\,6286S), a LIRG ($\log L_{\rm\,IR}/L_{\odot}$=11.36, \citealp{Howell:2010uq}) located at z=0.018349 (i.e., a luminosity distance of $d_{\rm\,L}=76.1$\,Mpc), which was not previously detected above 10\,keV \citep{Koss:2013pi}. The source has a star-formation rate (SFR) of 41.3\,$M_{\odot}\rm\,yr^{-1}$ \citep{Howell:2010uq}, is in an early merging stage (i.e.,  stage {\it B} or 2, following the classification of \citealp{Stierwalt:2013ff}), and is interacting with the galaxy NGC\,6285 (NGC\,6286N), located at a distance of $\sim 1.5$\,arcmin ($\sim$33\,kpc, projected distance, see Fig.\,\ref{fig:opticalimage} and panel four of Fig.\,\ref{fig:images}). The source is also known to host a OH megamaser \citep{Baan:1998ys}. The only previous X-ray study of this source, carried out using {\it XMM-Newton} observations, did not find any evidence of an AGN \citep{Brightman:2011oq}. The {\it XMM-Newton} can in fact be well represented by a model taking into account only a collisionally-ionized plasma and an unabsorbed power-law component, representing thermal emission from the starburst and X-ray radiation produced by X-ray binaries, respectively. Possible evidence of very faint AGN activity has been found studying the near-IR to radio spectral energy distribution (SED) \citep{Vega:2008kh}, and could be inferred by the detection of [Ne\,V] lines, although the detection of these features has been questioned by \cite{Inami:2013il}, and due to their weakness they might also be produced by a young starburst.

The paper is structured as follows. In $\S$\,\ref{sect:xrayobs_red} and $\S$\,\ref{Sect:radio} we present the X-ray and radio data used and describe the data reduction procedures, in $\S$\,\ref{sect:specAnalysis} we report on the X-ray spectral analysis of NGC\,6286, in $\S$\,\ref{sect:discussion} we discuss our results by taking into account the multi-wavelength properties of NGC\,6286, and in $\S$\,\ref{section:summary} we summarise the main results of our work. Throughout the paper we adopt standard cosmological parameters ($H_{0}=70\rm\,km\,s^{-1}\,Mpc^{-1}$, $\Omega_{\mathrm{m}}=0.3$, $\Omega_{\Lambda}=0.7$).

\section{X-ray observations and data reduction}\label{sect:xrayobs_red}

\subsection{{\it NuSTAR}}
{\it NuSTAR} observed NGC\,6286 on UT 2015 May 29 for 17.5\,ks. We processed the data using the {\it NuSTAR} Data Analysis Software \textsc{nustardas}\,v1.4.1 within HEASOFT\,v6.16, adopting the latest calibration files \citep{Madsen:2015uq}. The source is clearly detected in the 3--24\,keV image (panel three of Fig.\,\ref{fig:images}). For both focal plane modules (FPMA and FPMB) we extracted source and background spectra and light-curves with the \textsc{nuproducts} task. A circular region of 45\,arcsec was used for the source\footnote{In the 3--24\,keV band for photon indices $\Gamma=0.6-1.8$ this aperture encloses $\sim 65\%$ of the full PSF energy \citep{Lansbury:2015dq}.}, while the background was extracted from an annulus centred on the X-ray source, with inner and outer radii of 90 and 150 arcsec, respectively. The 3--10 and 10--50\,keV light-curves of the sources do not show any evidence of flux variability.

\subsection{{\it XMM-Newton}}

Two {\it XMM-Newton} \citep{Jansen:2001vn} observations of NGC\,6286 (ID: 0203390701 and 0203391201; PI: Maiolino) were carried out on UT 2005 May 7, with exposures of 20.8 and 8.9\,ks. Both PN \citep{Struder:2001uq} and MOS \citep{Turner:2001fk} data were analysed by reducing first the observation data files (ODFs) using the {\it XMM-Newton} Standard Analysis Software (SAS) version 12.0.1 \citep{Gabriel:2004fk}, and then the raw PN and MOS data files using the \textsc{epchain} and \textsc{emchain} tasks, respectively.  In order to filter the observations for periods of high background activity we analyzed the EPIC/PN and MOS background light curves in the 10--12 keV band and above 10\,keV, respectively, and found that both observations show a significant background contamination. Observation 0203391201 was not used because the background flux dominates the whole observation (with an average count-rate of $6\rm\,ct\,s^{-1}$ and a minimum of $2\rm\,ct\,s^{-1}$). Observation 0203390701 showed less contamination, and we filtered the periods of high background activity using a threshold of $2\rm\,ct\,s^{-1}$ for both PN and MOS, which resulted in net exposure times of 2.3 and 4.7\,ks, respectively. For both cameras we extracted the spectrum of the source using a circular region of 20\,arcsec radius, while the background was extracted from a circular region of 40\,arcsec radius, located on the same CCD of the source and in a zone devoid of other sources. No significant flux variability is found in the 0.3--10\,keV band during the {\it XMM-Newton} observation.

\medskip
\medskip

\subsection{{\it Swift}/XRT}
The X-ray Telescope (XRT, 0.2--10\,keV; \citealp{Burrows:2005vn}) on board {\it Swift} \citep{Gehrels:2004dq} observed NGC\,6286 quasi-simultaneously with {\it NuSTAR} on UT 2015 May 29 for 2\,ks. XRT data were reduced using the \textsc{xrtpipeline v0.13.0} within HEASOFT\,v6.16.

\subsection{{\it Chandra}}
A {\it Chandra} \citep{Weisskopf:2000vn} ACIS-S \citep{Garmire:2003kx} observation of NGC\,6286 was carried out on UT 2009 September 18 (PI: Swartz) with an exposure of 14.2\,ks. The data reduction was performed following the standard procedure, using \textsc{CIAO} v.4.6. The data were reprocessed using \textsc{chandra\_repro}, and then the spectra were extracted using the \textsc{specextract} tool. 

The 0.2--2\,keV {\it Chandra} image shows clear evidence of extended emission (panel one of Fig.\,\ref{fig:images}). The 3--8\,keV image (panel two) shows instead only a point-like source, which does not appear in the 0.2--2\,keV image. This source is located at the center of the galaxy (see panel three) and could be associated with AGN emission. The spectra of these different regions are discussed in $\S$\,\ref{sect:specAnalysis_extended_nuclear}. 

In order to be consistent with the spectral extraction of {\it XMM-Newton} and {\it Swift}/XRT, which have a much lower spatial resolution than {\it Chandra}, the ACIS-S source spectrum used for the broad-band X-ray fitting was extracted from a circular region of 10.5\,arcsec radius. The background spectrum was extracted from a circular region of the same size on the same CCD, where no other source was detected.

\begin{figure}
\includegraphics[width=8.5cm]{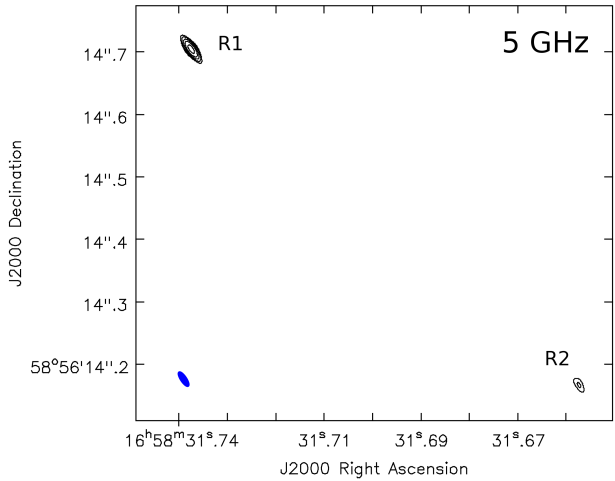}
\caption{EVN (Effelsberg, Westerbork and Lovell antennas) contour map of NGC\,6286 at 5\,GHz from UT 2005 June 13, imaged with a convolving beam of $8.39 \times 25.94$\,arcsec at 33.71\degr{} (North through East) using natural weighting.}
\label{fig:ep049c_cola172}
\end{figure}

\section{Radio observations and data reduction}\label{Sect:radio}
\cite{Parra:2010hc} reported VLA observations of NGC\,6286 at 4.8\,GHz, showing a compact morphology with a size of $0.25 \times 0.21$\,arcsec and a flux density of 15.24\,mJy. They also observed this galaxy with three of the most sensitive antennas (Effelsberg, Westerbork and Lovell) of the European very long baseline interferometry (VLBI) Network (EVN) at 5\,GHz, and detected fringes in each one of the baselines with amplitudes between 5.36--6.11\,mJy. We have extracted these observations from the archive and applied the pipelined calibration available. In Figure \ref{fig:ep049c_cola172} we show the contour map obtained using the cleaning algorithm within the Caltech program {\sc difmap} \citep{Shepherd:1995bs}. No proper flux density could be obtained with such a small array since the measurements are still subject of instrumental amplitude errors. We can however rely on the source structure, as there is enough information to determine phase closure. We find two milli-arcsec sources with $S/N>5$, one (R1) at RA$=16^{\mathrm{h}}58^{\mathrm{m}}31.7374^{\mathrm{s}}$, DEC = $+58\degr56\arcmin14.705\farcs$, and the other (R2) at RA=$16^{\mathrm{h}}58^{\mathrm{m}}31.6572^{\mathrm{s}}$, DEC$ = +58\degr56\arcmin14\farcs167$. These two sources are consistent with the 3--8\,keV core region (see panel two of Fig.\,\ref{fig:images} and $\S$\,\ref{sect:specAnalysis_extended_nuclear}).

We have also extracted and analysed the very long baseline array (VLBA) experiment BC196 observed at 8\,GHz on UT 2012 January 12. We used the NRAO Astronomical Image Processing System ({\sc aips}) to reduce the data, following standard procedures. We note that the source chosen as phase reference (J1651$+$5805) is not detected in this experiment, and constraints for it are also not available in the VLBA Calibrator search engine at NRAO. We have resorted to the use of another nearby calibrator (J1656$+$6012, at 2.22\degr) which was observed 2\,min before the NGC\,6286 scan. We found that there are no sources detected above $\sim 0.8$\,mJy/beam ($3 \times $r.m.s.) in the VLBA observations convolved with a $3.16 \times 0.94$\,arcsec at 29.31\degr{} beam. If any of the sources detected with
the EVN is the AGN core, we would expect a similar peak intensity measured in a baseline with comparable length as that from Ef-Jb or Ef-Wb baselines. The fact that we do not detect any source in the VLBA observations leaves two possible explanations: i) the sources are variable and the VLBA observations are not sensitive enough; ii) the sources are resolved at resolutions better than $\sim 3$\,arcsec. Although the VLBA array includes three times as much antennas as the EVN small array, it also observed the target for only 1/3 of the time with respect to the EVN, and using antennas $\sim$3--100 times smaller than those in the small EVN array. We made the exercise of producing an image with similar uv-range for both EVN and VLBA observations. The obtained uv-coverages result in the VLBA being sensitive to emission close to perpendicular to the structure we detected with the EVN (at an inclination $\sim 50\degr$), and since there is no emission in such orientation, the VLBA cannot detect any structure, unlike the EVN. Threfore, in order to better constrain the radio emission of NGC\,6286, further VLBI observations covering proper hour angles at high sensitivity are needed.

\begin{figure*}[t!]
\centering
\begin{minipage}[!b]{.45\textwidth}
\centering
\includegraphics[width=8.3cm]{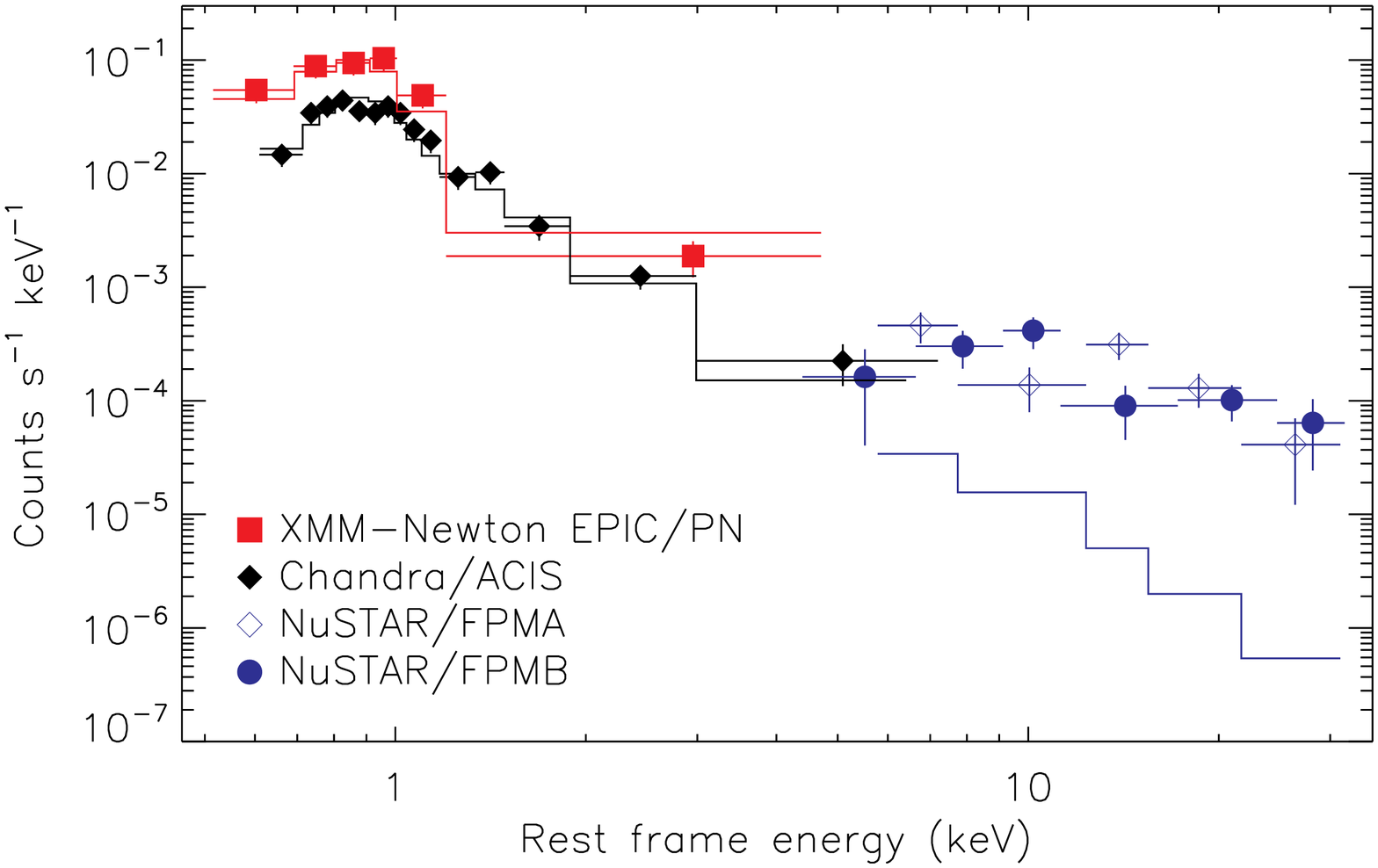}\end{minipage}
\begin{minipage}[!b]{.48\textwidth}
\centering
\includegraphics[width=8.3cm]{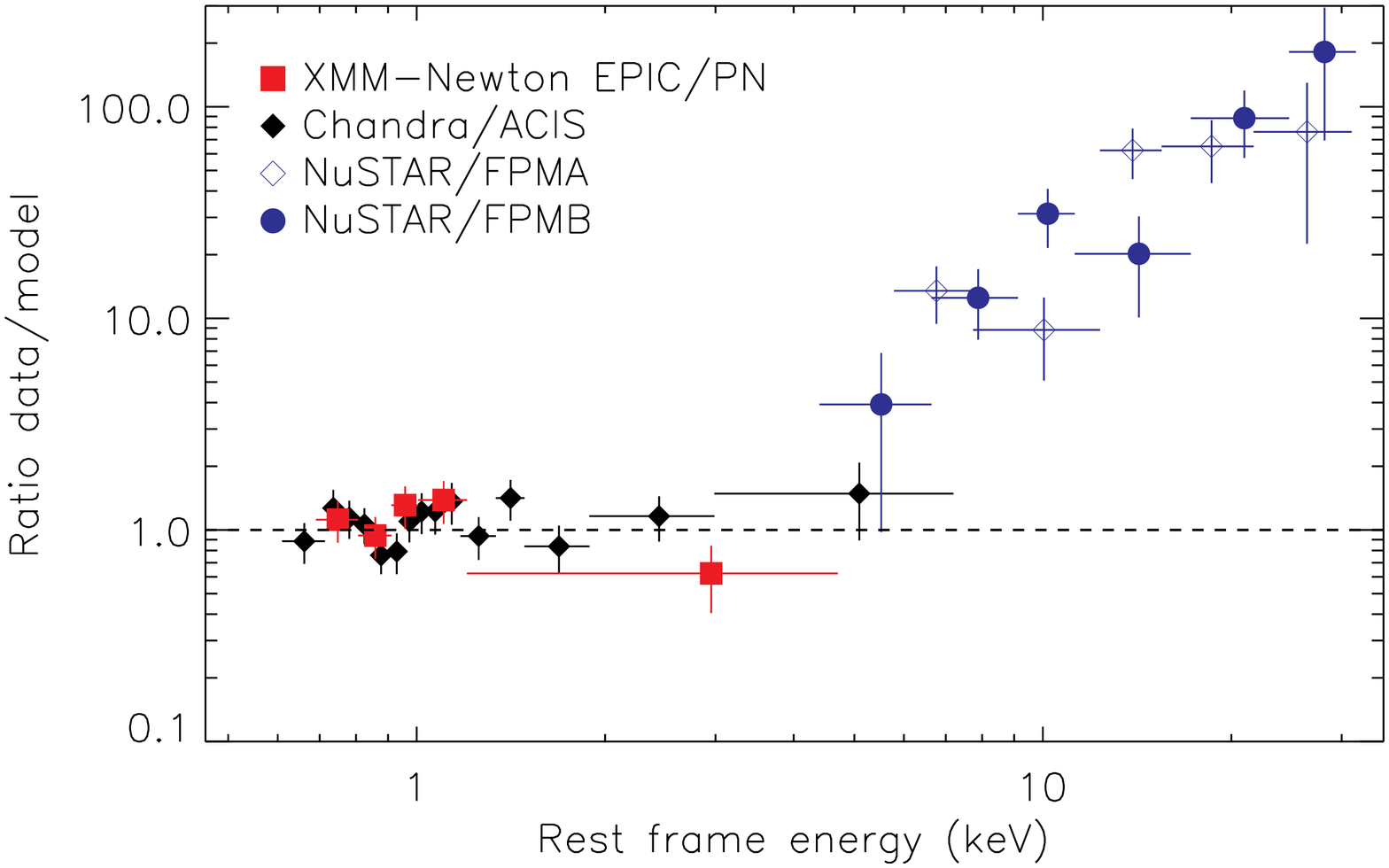}\end{minipage}
 \begin{minipage}[t]{\textwidth}
  \caption{{\it Left panel:} {\it XMM-Newton} EPIC/PN, {\it Chandra} ACIS-S (using a 10.5 arcsec extraction radius) and {\it NuSTAR} FPMA/FPMB spectra of NGC\,6286. The continuous lines represent the model used in \citet{Brightman:2011oq} (\textsc{apec+zpowerlaw}) to reproduce the 0.3--10\,keV spectrum of the source. {\it Right panel:} ratio between the data and the model. The plots clearly show the importance of hard X-ray coverage to fully understand the characteristics of the X-ray emission.  }
\label{fig:modelbn11}
 \end{minipage}
\end{figure*}

\section{X-ray spectral analysis}\label{sect:specAnalysis}
The X-ray spectral analysis was performed within \textsc{xspec}\,v.12.8.2 \citep{Arnaud:1996kx}. Galactic absorption in the direction of the source ($N_{\rm\,H}^{\rm\,G}= 1.8\times 10^{20}\rm\,cm^{-2}$, \citealp{Kalberla:2005fk}) was taken into account by adding photoelectric absorption (\textsc{TBabs} in \textsc{xspec}, \citealp{Wilms:2000vn}). Abundances were set to the solar value. Spectra were rebinned to have at least 20\,counts per bin in order to use $\chi^2$ statistics, unless reported otherwise.

In the following we first present the X-ray spectral analysis of the extended and nuclear emission revealed by {\it Chandra} ($\S$\,\ref{sect:specAnalysis_extended_nuclear}) and then discuss the spatially integrated broad-band X-ray emission ($\S$\,\ref{sect:allxray}) considering all observations available.

\subsection{Extended and nuclear emission}\label{sect:specAnalysis_extended_nuclear}

The diffuse soft X-ray emission detected by {\it Chandra} has an angular size of $\sim 12$\,arcsec, which at the distance of the source corresponds to $\sim 4.4$\,kpc. This diffuse emission might either be related to thermal plasma in a star-forming region, to X-ray binaries, or to shocks created by the interaction between outflows from the AGN and the galactic medium.
To analyse the diffuse and nuclear emission we extracted the spectra of the four regions shown in panel one of Fig.\,\ref{fig:images}. Besides the 3--8\,keV core, in order to study how the extended emission varies, we arbitrarily selected three regions (A, B and C) where most of the 0.2-2\,keV photons were detected. Due to the low number of counts we rebinned the spectra to have at least 1 count per bin, and used Cash statistics \citep{Cash:1979fk} to fit the data. In the following we discuss the spectral properties of the Core, and the regions A, B and C. 
\smallskip

\noindent{\bf Core.} The spectrum of the core was extracted from a circular region of radius 1.5\,arcsec centred on the peak of the 3--8\,keV emission. Ignoring the data below 1.2\,keV to avoid contamination from the diffuse soft X-ray emission, and fitting with a power-law model (\textsc{tbabs$_{\rm\,Gal}\times$zpowerlaw} in \textsc{xspec}) we obtain a photon index of $\Gamma=-0.17_{-1.03}^{+1.01}$. This low value is indicative of heavy absorption in the nuclear region. Fitting the X-ray spectrum using the whole energy range with a model that includes also a collisionally ionized plasma model (\textsc{tbabs$_{\rm\,Gal}\times$zpowerlaw+apec}) we obtain C-stat/DOF=20.9/21, $\Gamma\leq -0.03$ and a plasma temperature of $kT=0.99_{-0.35}^{+0.28}$\,keV. The 3--8\,keV core coincides with the 1.4\,GHz radio emission measured by the VLA FIRST survey \citep{Becker:1995fk}.

\begin{figure*}[t!]
\centering
\begin{minipage}[!b]{.45\textwidth}
\centering
\includegraphics[width=8.5cm]{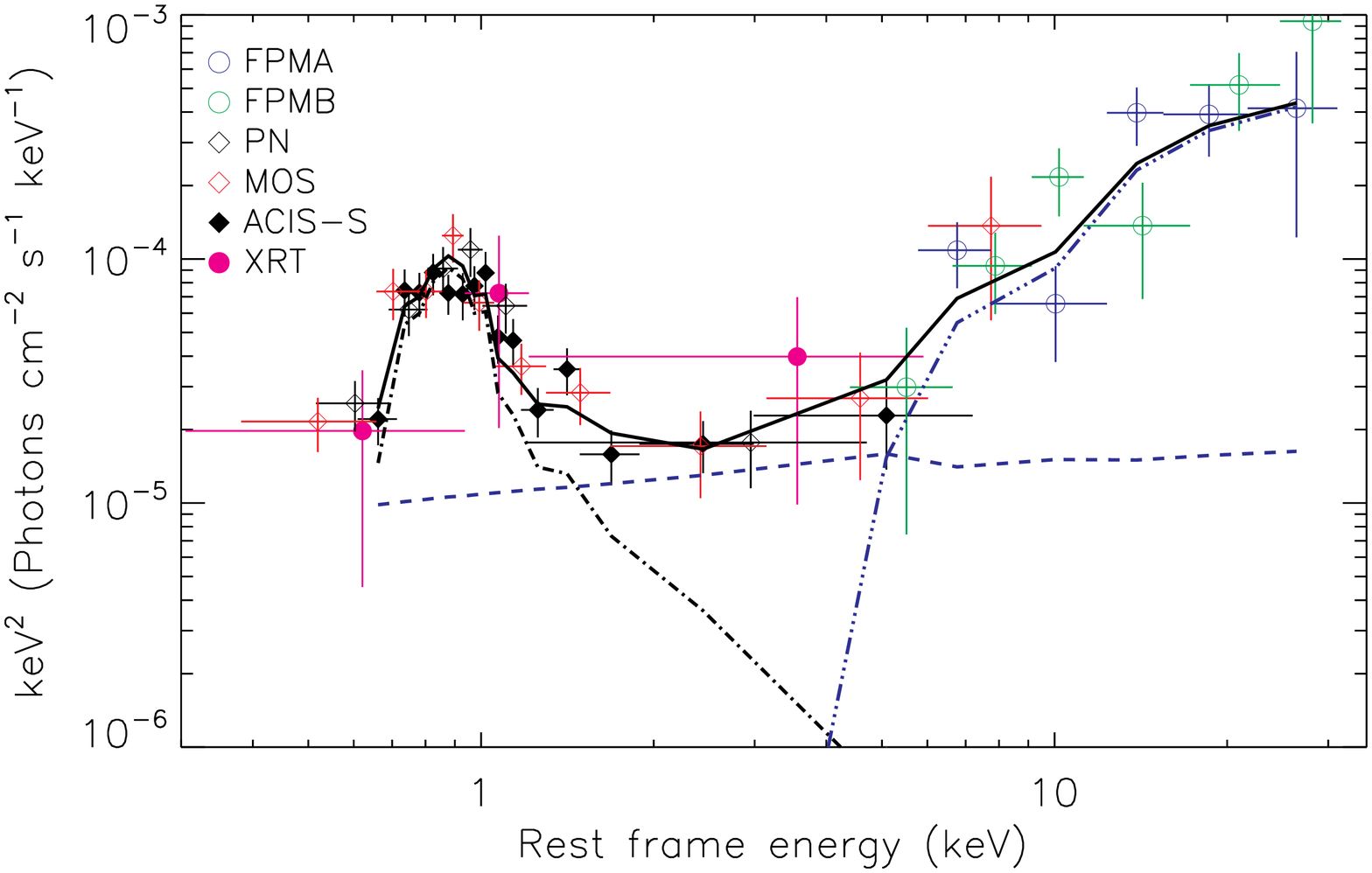}\end{minipage}
\begin{minipage}[!b]{.48\textwidth}
\centering
\includegraphics[width=8.5cm]{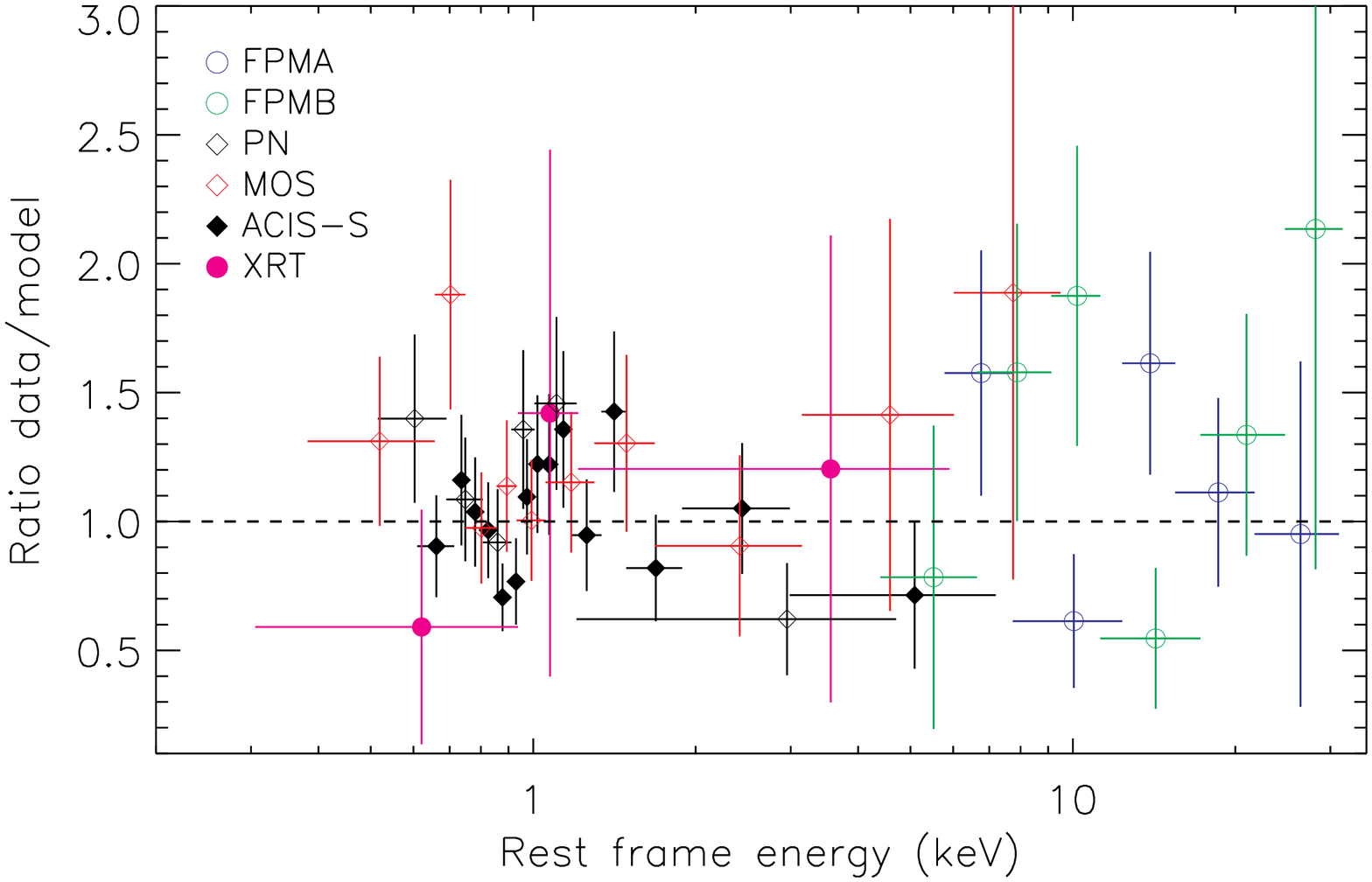}\end{minipage}
 \begin{minipage}[t]{\textwidth}
  \caption{{\it Left panel}: unfolded broad-band X-ray spectrum of NGC\,6286. The black continuous line represents the best fit to the data, while the dot-dashed line is the thermal plasma, the dashed line is the scattered emission and the dot-dot-dashed line is the \textsc{sphere} model. {\it Right panel}: ratio between the data and the model shown in the right panel.}
\label{fig:eeufs_ratio}
 \end{minipage}
\end{figure*}

\smallskip
\noindent{\bf Region A.} Fitting the spectrum with a collisionally ionized thermal plasma model (\textsc{tbabs$_{\rm\,Gal}\times$apec}) results in a good fit (C-stat/DOF=31.0/37), with $kT=0.91^{+0.10}_{-0.18}$\,keV. 
Applying a spectral model which reproduces a non-equilibrium plasma created in a shock (\textsc{pshock} in \textsc{xspec}) yields C-stat/DOF=31.4/36, a plasma temperature of $kT_{\rm\,S}=0.88^{+0.11}_{-0.13}$\,keV and an upper limit on the ionization timescale of $\tau_{\rm\,u}\geq 1.6 \times 10^{12}\rm\,s\,cm^{-3}$. 

\smallskip
\noindent{\bf Region B.} Using the thermal plasma model yields a rather poor fit (C-stat/DOF=48.0/34). This can be improved by adding photoelectric absorption (\textsc{tbabs$_{\rm\,Gal}\times$ztbabs$\times$apec}, C-stat/DOF=43.7/33), and would be consistent with the presence of larger absorption in the central part of the edge-on galaxy with respect to other regions. The shock plasma model fails to reproduce well the spectrum both without (C-stat/DOF=48.7/33) and with (C-stat/DOF=43.6/32) an absorption component.

\smallskip
\noindent{\bf Region C.} A thermal plasma model with $kT=1.23^{+0.37}_{-0.35}$\,keV yields a good fit (C-stat/DOF=34/34), while a shock plasma model cannot reproduce well the data (C-stat/DOF=45.1/33), and results in $kT_{\rm\,s}=1.09^{+0.73}_{-0.38}$\,keV and $\tau_{\rm\,u}\leq 1.4\times 10^8 \rm\,s\,cm^{-3}$.

\subsection{Spatially integrated X-ray emission}\label{sect:allxray}

The {\it XMM-Newton} EPIC spectrum of NGC\,6286 was analysed by \cite{Brightman:2011oq}, who found that it could be well represented by an unabsorbed power-law continuum plus a thermal plasma\footnote{\textsc{zpowerlaw+apec} in \textsc{xspec}}, with the photon index fixed to $\Gamma=1.9$. This is in disagreement with the hard ($\Gamma=-0.17_{-1.03}^{+1.01}$, see $\S$\,\ref{sect:specAnalysis_extended_nuclear}) 1--8\,keV spectrum of the 3--8\,keV core. Fitting the {\it NuSTAR} FPMA/FPMB data with a simple power-law model we also find a very flat continuum ($\Gamma=0.49^{+0.46}_{-0.41}$). The low values of the photon index obtained in the 3--8\,keV and 3-30\,keV bands could indicate that the X-ray emission is highly absorbed.

While the model used by \cite{Brightman:2011oq} can reproduce well the {\it XMM-Newton} and the spatially integrated {\it Chandra} spectra, it severely under predicts the hard X-ray flux inferred by {\it NuSTAR}, as illustrated in Fig.\,\ref{fig:modelbn11}. This might be related either to heavy obscuration of the X-ray source or to flux variability between {\it XMM-Newton} and {\it NuSTAR} observations, although variability would not be able to explain the very flat hard X-ray spectrum. The {\it Swift}/XRT observation allows us to constrain the flux level below 10\,keV band at the time of the {\it NuSTAR} observations. We find that the {\it Swift}/XRT 0.3--2 and 2--5\,keV fluxes are consistent with that inferred by {\it Chandra} and {\it XMM-Newton} EPIC/PN observations (see Table\,\ref{tab:X-rayflux}), which implies the lack of significant variability between the different observations. To further test the variability scenario we fitted {\it NuSTAR} and the spatially integrated {\it Chandra} spectra with a model that consists of a power-law plus a thermal plasma [\textsc{tbabs$_{\rm\,Gal}\times$(apec + power law)}], allowing for different normalisations of the power-law continuum to vary (fixing $\Gamma=1.9$). We found that the model cannot reproduce well the spectra ($\chi^2$/DOF=28.9/20), with the fit\footnote{the ratio of the power-law normalisations is $\simeq 4$.} showing clear residuals between 10 and 30\,keV. This rules out variability as a likely explanation for the large ratio between the 10--50\,keV and 2--10\,keV fluxes.

\begin{table}
\begin{center}
\caption[]{Observed X-ray fluxes}
\label{tab:X-rayflux}
\begin{tabular}{lcc}
\noalign{\smallskip}
\hline \hline \noalign{\smallskip}
\noalign{\smallskip}
	  	Facility		&   \multicolumn{2}{c}{Flux} 	\\
\noalign{\smallskip}
\hline
\noalign{\smallskip}
	  			&    0.3--2\,keV   & 2--5\,keV 	\\
\noalign{\smallskip}
	  		&   [\scriptsize{$10^{-14}\rm\,erg\,s^{-1}\,cm^{-2}$}]  &   [\scriptsize{$10^{-14}\rm\,erg\,s^{-1}\,cm^{-2}$}]  	\\
\noalign{\smallskip}
\hline \noalign{\smallskip}
	{\it Swift}/XRT		&	$8.4_{-2.9}^{+4.1}$ & $2.0_{-1.0}^{+0.8}$	\\
\noalign{\smallskip}
	{\it Chandra}		&	$8.8_{-0.6}^{+0.6}$ &  $2.0_{-0.4}^{+0.3}$		\\
\noalign{\smallskip}
	{\it XMM-Newton}	&	$9.5_{-1.1}^{+0.8}$ & $2.2_{-0.4}^{+0.2}$	\\
\noalign{\smallskip}
\hline
\noalign{\smallskip}

\end{tabular}
\end{center}
\end{table}

In the following we report the results obtained by adopting several different X-ray spectral models to infer the properties of the AGN in NGC\,6286. In order to reduce the possible degeneracies in the models, we constrained the average properties of the diffuse soft X-ray emission. To do this we first extracted the {\it Chandra} X-ray spectrum of the diffuse emission by excluding from the circular region of 10.5\,arcsec a circle of 1.5\,arcsec centred on the 3--8\,keV core. We then fitted the spectrum with a model that includes i) a thermal plasma and ii) a power-law component ($\Gamma=1.9$) to take into account the scattered emission. We obtained a normalization of the power law $n_{\rm\,po}^{\rm\,scatt}=(1.03\pm0.37)\times 10^{-5}\rm\,ph\,keV^{-1}\,cm^{-2}\,s^{-1}$, and a temperature and normalization of the thermal plasma of $kT=0.77^{+0.07}_{-0.08}$\,keV and $n_{\rm\,apec}=(2.02\pm0.33)\times 10^{-5}\rm\,ph\,keV^{-1}\,cm^{-2}\,s^{-1}$, respectively. In all the spectral models reported below we set $n_{\rm\,po}^{\rm\,scatt}$, $kT$ and $n_{\rm\,apec}$ to the values obtained for the diffuse emission, and allow them to vary only within their 90\% uncertainties.

\subsubsection{PEXRAV}

To infer the value of the line-of-sight column density ($N_{\rm\,H}$) we fitted the joint {\it Swift}/XRT, {\it Chandra} ACIS-S, {\it XMM-Newton} EPIC/PN and MOS, and {\it NuSTAR} FPMA and FPMB data with a model that consists of: a) an absorbed power-law with a photon index fixed to $\Gamma=1.9$, consistent with the average value of AGN (e.g., \citealp{Nandra:1994kl,Piconcelli:2005tg,Ricci:2011oq}), b) unabsorbed reprocessed X-ray emission from a slab, c) a Gaussian to reproduce the fluorescent Fe K$\alpha$ emission line (with the rest-frame energy fixed to $E_{\rm\,K\alpha}=6.4$\,keV), d) a second power-law to reproduce the scattered component, and e) emission from a collisionally ionized plasma. To reproduce the effect of obscuration we included both Compton scattering and photoelectric absorption. Reprocessed X-ray emission (excluding fluorescent lines) was taken into account using the \textsc{pexrav} model \citep{Magdziarz:1995pi}. The fraction of scattered flux ($f_{\rm\,scatt}$) is calculated as the ratio between the normalization at 1\,keV of the primary power law ($n_{\rm\,po}$) and $n_{\rm\,po}^{\rm\,scatt}$. The width of the Gaussian line was fixed to $\sigma=40$\,eV, consistent with the results obtained by {\it Chandra}/HETG studies (e.g., \citealp{Shu:2010zr}).  An Fe K$\alpha$ line at 6.4\,keV is usually found in the X-ray spectrum of AGN (e.g., \citealp{Nandra:1994fk}, \citealp{Shu:2010zr}, \citealp{Ricci:2014ys}), and is believed to originate in the material surrounding the SMBH (e.g., \citealp{Ricci:2014zr}, \citealp{Gandhi:2015lh} and references therein). In \textsc{xspec} our model is:

\smallskip
\noindent\textsc{tbabs$_{\rm\,Gal}$(ztbabs$\times$cabs$\times$zpowerlaw + pexrav + zgauss + apec + zpowerlaw)}. 
\smallskip

\noindent The model yields a good fit ($\chi^2$/DOF=47.2/44) and results in a column density consistent with border-line Compton-thick obscuration ($N_{\rm\,H}=1.32^{+0.82}_{-0.54}\times 10^{24}\rm\,cm^{-2}$). Due to the low signal-to-noise ratio, the Fe K$\alpha$ is not spectrally resolved, and only an upper limit of its equivalent width was obtained ($EW\leq 2318$\,eV), which is consistent with heavy obscuration.

\begin{figure}[t!]
\centering
\includegraphics[width=8.5cm]{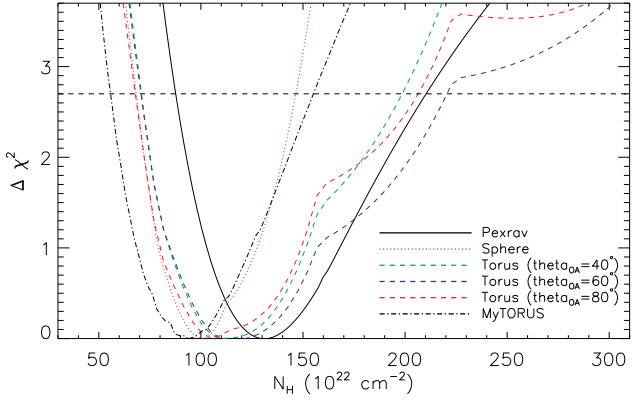}
  \caption{Value of $\Delta \chi^2=\chi^2-\chi^2_{\rm best}$ (where $\chi^2_{\rm best}$ is the minimum value of the $\chi^2$) versus the column density for the different X-ray spectral models discussed in $\S$\,\ref{sect:allxray}. The horizontal dashed line represents $\Delta \chi^2=2.7$. The plot shows that NGC\,6286 is heavily obscured, with $N_{\rm\,H}$ consistent with the source being CT for the five models considered.}
\label{fig:dchiNH}
\end{figure}

\subsubsection{TORUS}

To further study the absorbing material we used the \textsc{torus} model developed by \cite{Brightman:2011oq}, which considers reprocessed and absorbed X-ray emission from a spherical-toroidal structure. In this model the line-of-sight column density is independent of the inclination angle, which we fixed to the maximum value permitted ($\theta_{\rm\,i}=87.1^{\circ}$). Similarly to what was done for \textsc{pexrav}, we added to the model a power law, to take into account the scattered emission, and a collisionally ionized plasma model. In \textsc{xspec}, the model is

\smallskip
\noindent\textsc{tbabs$_{\rm\,Gal}$(atable\{torus1006.fits\} + apec + zpowerlaw)}. 
\smallskip

\noindent We fixed $\Gamma=1.9$ and tested several values of the half-opening angle of the torus ($\theta_{\rm\,OA}=40^{\circ}, 60^{\circ}, 80^{\circ}$). The three models are statistically indistinguishable, and in all cases we obtained good fits. For the three values of $\theta_{\rm\,OA}$ the column densities are consistent within the uncertainties with CT absorption. 

\subsubsection{SPHERE}

To test the scenario in which the X-ray source is fully covered by the obscuring material we applied the \textsc{sphere} model \citep{Brightman:2011oq}, using the same setting as for the \textsc{torus} model:

\smallskip
\noindent \textsc{tbabs$_{\rm\,Gal}$(atable\{sphere0708.fits\} + apec + zpowerlaw)}.  
\smallskip

\noindent This model provides a good fit (Fig.\,\ref{fig:eeufs_ratio}), and confirms the presence of heavy obscuration ($N_{\rm\,H}=9.8^{+4.6}_{-3.5}\times 10^{23}\rm\,cm^{-2}$).

\smallskip
\subsubsection{MYTORUS}

Next we applied the \textsc{MYTorus} model \citep{Murphy:2009ly}, which considers absorbed and reprocessed X-ray emission from a smooth torus with $\theta_{\mathrm{OA}} = 60^{\circ}$, and can be used for spectral fitting as a combination of three additive and exponential table models: the zeroth-order continuum (\textsc{mytorusZ}), the scattered continuum (\textsc{mytorusS}) and a component containing the fluorescent emission lines (\textsc{mytorusL}). We used the {\it decoupled} version of \textsc{MYTorus} \citep{Yaqoob:2012uq}. This was done by: i) allowing the values of the column density of the absorbing [$N_{\rm\,H}^{\rm\,T}(Z)$] and reprocessing [$N_{\rm\,H}^{\rm\,T}(S,L)$] material to have different values; ii) fixing the inclination angle of \textsc{mytorusL} and \textsc{mytorusS} to $\theta_{\rm\,i}(S,L)=0^{\circ}$, and that of \textsc{mytorusZ} to $\theta_{\rm\,i}(Z)=90^{\circ}$; iii) adding a second scattered component with $\theta_{\rm\,i}(S,L)=90^{\circ}$; iv) leaving the normalizations of the transmitted and scattered component ($n_{\rm\,po}$ and $n_{\rm\,refl}$) free to vary. To this model we added a scattered component and thermal emission. In \textsc{xspec} the model is:

\smallskip
\textsc{tbabs$_{\rm\,Gal}\times$\{mytorusZ($90^{\circ}$)$\times$ zpowerlaw + mytorusS($0^{\circ}$) + mytorusS($90^{\circ}$) + gsmooth[mytorusL($0^{\circ}$) + mytorusL($90^{\circ}$)] + apec + zpowerlaw\}}. 
\smallskip

\noindent Due to the low signal-to-noise ratio of the spectrum we could not constrain the different values of $N_{\rm\,H}^{\rm\,T}(S,L)$ and $N_{\rm\,H}^{\rm\,T}(Z)$, so their values were tied. The same was done for the normalizations of the scattered and transmitted components, while the photon index was left free to vary. This model also yields a good fit and results in a line-of-sight column density consistent with heavy obscuration ($N_{\rm\,H}=8.8^{+5.1}_{-3.8}\times 10^{23}\rm\,cm^{-2}$).

\smallskip
\smallskip
\smallskip
The parameters obtained from the spectral analysis are reported in Table\,\ref{tab:X-rayresults}, while in Figure\,\ref{fig:dchiNH} we show the values of $\Delta \chi^2$ versus $N_{\rm\,H}$ for the models described above. Depending on the X-ray spectral model adopted, the intrinsic (i.e. absorption and k-corrected) 2--10\,keV luminosity of NGC\,6286 is $3-20\times 10^{41}\rm\,erg\,s^{-1}$.

\begin{table*}
\begin{center}
\caption[]{List of IR and X-ray tracers of AGN activity commonly used for U/LIRGs.}
\label{tab:AGN_tracers}
\begin{tabular}{lccccc}
\noalign{\smallskip}
\hline \hline \noalign{\smallskip}
%
	  	(1)	& (2)	& (3) & (4)  & (5)  & (6)  	\\
\noalign{\smallskip}
	  	Indicator	& NGC\,6286	& Reference & Mean GOALS  & Threshold  & AGN  	\\
%
\noalign{\smallskip}
\hline \noalign{\smallskip}
[Ne\,V]\,$14.32\mu$m	[$10^{-17}\rm\,W\,m^{-2}$] 		&	 $0.33\pm0.11$ 		& \cite{Dudik:2009cq} & 2.27$^{\rm A}$  &\nodata & ?	\\
\noalign{\smallskip}
[Ne\,V]\,$24.32\mu$m 	[$10^{-17}\rm\,W\,m^{-2}$]	 	&	$0.99\pm0.20$		& \cite{Dudik:2009cq} & \nodata & \nodata & ?	\\
\noalign{\smallskip}
[Ne\,V]/[Ne\,II]  				&	0.02 		& \cite{Dudik:2009cq} &  0.07$^{\rm B}$ & $\geq 0.1$$^{\rm C}$	 & \xmark  \\
\noalign{\smallskip} 
[OIV]/[Ne\,II]	&	 0.05		& \cite{Dudik:2009cq}  & 0.03$^{\rm D}$/0.24$^{\rm E}$ & $\geq 1.75$$^{\rm F}$ &  \xmark	\\
\noalign{\smallskip}
$\Gamma_{2.5-5\mu\rm m}$	&$-0.1$	 		& \cite{Imanishi:2010uq} & \nodata & $\geq 1$$^{\rm G}$	  &  \xmark  \\
\noalign{\smallskip}
$F_{\nu}(30\mu\mathrm{m})/F_{\nu}(15\mu\rm m)$	&	 5.97		& \cite{Stierwalt:2013ff}  & $8^{+2}_{-1.5}$$^{\rm H}$ & \nodata &\xmark 	\\
\noalign{\smallskip}
$EW(\rm PAH\,\,3.3\mu m) [nm]$	&	48 		&\cite{Imanishi:2010uq}  &  \nodata &	$<40$$^{\rm G}$ & \xmark  \\
\noalign{\smallskip}
$EW(\rm PAH\,\,6.2\mu\rm\,m)$ [$\mu$m]	&	0.59 		& \cite{Stierwalt:2013ff}  & $0.55$$^{\rm H}$ & $\leq 0.3^{\rm C}$ &\xmark 	\\
\noalign{\smallskip}
$\tau_{3.1\mu\rm m}$ (3.1$\mu\rm m$ H$_2$O ice)	&	 ND		& \cite{Imanishi:2010uq} & \nodata &  $>0.3$$^{\rm G}$	& \xmark \\
\noalign{\smallskip}
$\tau_{3.4\mu\rm m}$ (3.4$\mu\rm m$ bare carbonaceous)	&	ND 		& \cite{Imanishi:2010uq} & \nodata  & 	$>0.2$$^{\rm G}$ & \xmark  \\
\noalign{\smallskip}
 $\tau_{9.7\mu\rm m}$ 	&	$-0.40$ 		& \cite{Stierwalt:2013ff} & $-0.35$$^{\rm H}$ & \nodata  &	\xmark \\
\noalign{\smallskip}
{\it Chandra} hardness ratio	&	$-0.85\pm0.07$ 		& This work & $-0.56$$^{\rm I}$ & $>-0.3$$^{\rm J}$ &\xmark 	\\
\noalign{\smallskip}
Observed $\log L_{\rm\,2-10}$ [$\rm\,erg\,s^{-1}$]	&	 40.80		& This work  & 41.3$^{\rm K}$ & $>42$$^{\rm L}$	&  \xmark  \\
\noalign{\smallskip}
Radio/FIR flux ratio (q)	&	2 		&  \cite{U:2012fc}  & $2.41\pm 0.29$$^{\rm M}$ & $<1.64$$^{\rm N}$ & \xmark 	\\
\noalign{\smallskip}
%
\hline
\noalign{\smallskip}
\multicolumn{6}{l}{The table lists (1) the indicator used, (2) the value and (3) reference for NGC\,6286, (4) the mean value for the  }\\
\multicolumn{6}{l}{GOALS sample, (5) the threshold used to infer the presence of AGN, and (6) whether an AGN was found or not.  }\\
\multicolumn{6}{l}{{\bf Notes}. ND: not detected; $^{\rm A}$ median of the 43 detections (18\% of the sample) from \cite{Petric:2011zt};}\\
\multicolumn{6}{l}{$^{\rm B}$ median \citep{Petric:2011zt}; $^{\rm C}$ threshold used by \cite{Inami:2013il} to establish a significant contribution }\\
\multicolumn{6}{l}{of the AGN to the MIR flux. $^{\rm D}$ median and $^{\rm E}$ mean from \cite{Petric:2011zt}; $^{\rm F}$ Value indicating if the AGN }\\
\multicolumn{6}{l}{contributes to more than $50\%$ of the MIR flux \citep{Petric:2011zt}; $^{\rm G}$ value used by \cite{Imanishi:2010uq}}\\
\multicolumn{6}{l}{$^{\rm H}$ mean value for objects in the same merger stage (B) as NGC\,6286 \citep{Stierwalt:2013ff}, the $30\mu\rm m/15\mu\rm m$}\\
\multicolumn{6}{l}{flux density ratio of NGC\,6286  is only marginally lower than the average value for the B merger stage, and }\\
\multicolumn{6}{l}{has a value consistent with 63\% of GOALS LIRGs; $^{\rm I}$ median of the 44 objects reported  in \cite{Iwasawa:2011fk};}\\
\multicolumn{6}{l}{  $^{\rm J}$ value used to establish the presence of an AGN and $^{\rm K}$ median value \citep{Iwasawa:2011fk}; $^{\rm L}$ Values  }\\
\multicolumn{6}{l}{commonly used to separate AGN from starbursts in the 2--10\,keV band (e.g., \citealp{Szokoly:2004uq}, }\\
\multicolumn{6}{l}{ \citealp{Kartaltepe:2010qf}); $^{\rm M}$ mean obtained for the 64 objects studied by \cite{U:2012fc}; $^{\rm N}$ threshold for }\\
\multicolumn{6}{l}{ radio-excess defined by \cite{Yun:2001rq}. }\\

\end{tabular}
\end{center}
\end{table*}

\section{Discussion}\label{sect:discussion}

The X-ray spectral analysis of NGC\,6286 reported above clearly shows that the accreting SMBH is heavily obscured, possibly by CT material (see Fig.\,\ref{fig:dchiNH}). The very flat continuum found by both {\it Chandra} (for the hard X-ray core) and {\it NuSTAR}, together with the fact that the 1.4\,GHz emission coincides with the 3--8\,keV {\it Chandra} point-source (Panel\,2 of Fig.\,\ref{fig:images}) confirms the presence of a heavily obscured AGN. While the buried AGN in NGC\,6286 could be easily identified at hard X-rays, several other diagnostics failed to detect it because of its low-luminosity. In $\S$\,\ref{sect:tracers} we illustrate the most commonly adopted techniques to detect AGN in U/LIRGs, and discuss the case of NGC\,6286 by exploiting the wealth of multi-wavelength data available for the GOALS sample. In $\S$\,\ref{sect:IRlumAGN} we estimate the contribution of the AGN to the luminosity of NGC\,6286, while in $\S$\,\ref{sect:optradio} we discuss the optical and radio properties of the galaxy, comparing them to those of other similar LIRGs. Finally, in $\S$\,\ref{sect:obscuredAGNinLIRGs}, we discuss the presence of heavily obscured low-luminosity AGN in LIRGs.

\smallskip
\smallskip

\subsection{IR and X-ray tracers of AGN activity in U/LIRGs}\label{sect:tracers}

AGN in U/LIRGs can be identified in the IR by several means: i) with the detection of high-excitation MIR emission lines (e.g., \citealp{Sturm:2002tw}), and in particular of [Ne\,V]\,$14.32\mu$m and [Ne\,V]\,$24.32\mu$m (e.g., \citealp{Weedman:2005cr}, \citealp{Satyapal:2008uq}, \citealp{Goulding:2009dq}); ii) using the ratios of high-to-low ionization fine-structure emission lines (e.g., [Ne\,V]\,$14.32\mu$m/[Ne\,II]\,$12.8\mu$m and  [OIV]\,$25.89\mu$m/[Ne\,II]\,12.8$\mu$m; e.g., \citealp{Lutz:1999jl}, \citealp{Petric:2011zt});
iii) with the EW of the PAH features, which tend to be lower in the presence of a bright AGN, since it can destroy PAH molecules (e.g., \citealp{Imanishi:2010uq});
iv) studying the slope of the 2.5--5$\mu$m continuum ($\Gamma_{2.5-5}$, e.g., \citealp{Imanishi:2010uq}) or the continuum $30\mu\rm m/15\mu\rm m$ flux density ratio (e.g., \citealp{Stierwalt:2013ff}), which tend to be red in the presence of an AGN;
v) using the depth of absorption features (e.g., \citealp{Imanishi:2000fu,Risaliti:2006kx,Georgantopoulos:2011uq}), with large depths pointing towards AGN obscured by dust;
and/or vi) from deviations of the well known correlation between the far-IR (FIR) and the radio luminosity (\citealp{Helou:1985ss}, \citealp{Condon:1991qc}, \citealp{Condon:1992mw}), using the radio-FIR flux ratio $q$ (e.g., \citealp{Yun:2001rq}).
We find that all these proxies (Table\,\ref{tab:AGN_tracers}) fail to detect the AGN in NGC\,6286, with the exception of the Ne\,V lines. These lines can however also be produced by a young starburst with a large population of Wolf-Rayet and O stars (e.g., \citealp{Abel:2008kx}), so their detection does not always provide conclusive evidence of an AGN. This is especially true for NGC\,6286, since the Ne\,V lines are weak [$\log (L_{\rm\,[Ne\,V]}/\rm erg\,s^{-1})\sim 38.8$]. Moreover, the detection of the [Ne\,V] lines has been questioned by \cite{Inami:2013il}, who found [Ne\,V]\,$14.32\mu$m to be detected only in one of the two {\it Spitzer} observations available, while in both observations a feature possibly consistent with [Ne\,V]\,$24.32\mu$m were detected at $\sim 24.37\mu$m (Inami, private communication).
{\it Spitzer}/IRAC selection provides another important tool for identifying AGN (e.g., \citealp{Lacy:2004ly}, \citealp{Stern:2005ve}). Using the AGN selection criteria proposed by \cite{Donley:2012bd} (see Eq.1 and 2 in their paper), and considering the fluxes reported by \cite{U:2012fc}, we find that NGC\,6286 does not satisfy the conditions for the presence of an AGN. The fact that the IR proxies fail to identify the AGN emission in NGC\,6286 is due to the problematic identification of low-luminosity AGN with IR spectra dominated by the host. For example, in a low-luminosity AGN the silicate absorption feature would be diluted by the strong IR continuum of the host galaxy.

\cite{Iwasawa:2011fk} studied 44 LIRGs from the GOALS sample with {\it Chandra}, and assessed the presence of an AGN using the hardness ratio $HR\equiv (H-S)/(H+S)$, where $H$ and $S$ are the background-corrected counts in the 2--8 and 0.5--2\,keV ranges, respectively. Sources with $HR>-0.3$ are reported as candidate AGN. This value was chosen because ULIRGs which are known to host AGNs, such as Mrk\,231, Mrk\,273, and UGC\,5101, tend to cluster just above this limit \citep{Iwasawa:2009vn}. Considering the spatially-integrated X-ray flux NGC\,6286 has a hardness ratio $HR=-0.85\pm0.07$, which would not allow to infer the presence of an AGN. However, as discussed by \cite{Iwasawa:2011fk} this threshold could become less reliable for some CT AGN, since mostly reprocessed radiation is observed in the hard X-ray band. 
Another criteria commonly used to identify AGN is the observed 2--10\,keV X-ray luminosity. Using $\log (L_{2-10}/\rm\,erg\,s^{-1})>42$ as a threshold (e.g., \citealp{Szokoly:2004uq}, \citealp{Kartaltepe:2010qf}), one would also miss identifying NGC\,6286 as a buried AGN [$\log (L_{2-10}/\rm\,erg\,s^{-1})=40.80$].

Spectral decomposition (e.g., \citealp{Nardini:2008hs}, \citealp{Alonso-Herrero:2012ud}) is another powerful method to constrain the contribution of AGN to the multi-wavelength SED. \cite{Vega:2008kh} found that a pure starburst model fails to reproduce well the near-IR to radio SED of NGC\,6286, and a buried AGN accounting for 5\% of the IR luminosity is required by the data.
{An useful diagnostic of the presence of a heavily obscured AGN is the ratio between the MIR and the 2--10\,keV luminosities (e.g., \citealp{Alexander:2008kx,Rovilos:2014ys,Georgantopoulos:2011zr}). It has been shown indeed that for AGN the absorption-corrected 2--10\,keV and the 6 and 12\,$\mu$m luminosities are well correlated (e.g., \citealp{Gandhi:2009zr,Stern:2015ve,Asmus:2015ly}), so that deviations from the correlation might imply the presence of heavy obscuration. \cite{Vega:2008kh} report that at 6$\mu$m about $58$\% of the flux is produced by the AGN. This would imply that the ratio between the IR and observed X-ray AGN luminosity is very low: $L_{2-10}/L_{6\mu\rm m}\simeq 2.4\times 10^{-3}$. This value is consistent with undetected DOGs in the CDF-N \citep{Georgakakis:2010ly} and with other U/LIRGs \citep{Georgantopoulos:2011zr}, which is related to the fact that in U/LIRGs the IR emission is enhanced by strong star formation, leading to very low values of $L_{2-10}/L_{6\mu\rm m}$. Using the largest 2--10\,keV X-ray luminosity obtained in $\S$\,\ref{sect:specAnalysis} ($L_{2-10}\sim2\times 10^{42}\rm\,erg\,s^{-1}$) one would still find that $L_{2-10}/L_{6\mu\rm m}\sim 0.1$, a value lower than that expected from the $L_{2-10}-L_{6\mu\rm m}$ correlation. This might imply that the AGN contribution to the IR flux is significantly lower than that reported by \cite{Vega:2008kh} (see $\S$\ref{sect:IRlumAGN} and Fig.\,\ref{fig:lirtot_lx}).

\subsection{AGN contribution to the IR luminosity}\label{sect:IRlumAGN}

The IR luminosity of NGC\,6286 is $8.8\times 10^{44}\rm\,erg\,s^{-1}$, which would imply that, depending on the X-ray spectral model used, we obtain a ratio $L_{2-10}/L_{\rm\,IR}\simeq4\times10^{-4}-2.3\times10^{-3}$, significantly lower than the value expected from pure AGN (e.g., \citealp{Mullaney:2011mb}). Considering the observed 2--10\,keV luminosity, the ratio is $\log(L_{2-10}^{\rm\,obs}/L_{\rm\,IR})\simeq -4.14$, which is consistent with the average value found for the GOALS sample [$\log (L_{2-10}^{\rm\,obs}/L_{\rm\,IR})=-4.40\pm0.63$, \citealp{Iwasawa:2011fk}].

Using the relationship of \citet{Mullaney:2011mb}, it is possible to convert the 2--10\,keV luminosity into the expected IR luminosity emitted by the dust around the AGN:
\begin{equation}\label{eq:xtoIR}
\log L^{\rm\,AGN}_{\rm\,IR,\,43}=(0.53\pm0.26)+(1.11\pm0.07) \log L_{2-10,43}.
\end{equation}
In the above equation $L^{\rm\,AGN}_{\rm\,IR,\,43}$ and $L_{2-10,43}$ are the 8--1000$\mu$m and 2--10\,keV luminosities of the AGN in units of $10^{43}\rm\,erg\,s^{-1}$. Considering the range of values obtained for the 2--10\,keV intrinsic luminosity, the IR luminosity of the AGN is $\log (L^{\rm\,AGN}_{\rm\,IR}/\rm\,erg\,s^{-1})=41.91-42.75$. Comparing this to the IR luminosity of the system [$\log (L_{\rm\,IR}/\rm\,erg\,s^{-1})=44.96$] we find that the IR luminosity of the AGN is between 0.1 and 0.6\% of the total IR luminosity. 
This value is in disagreement with that obtained by \cite{Vega:2008kh} using spectral decomposition, who found that the contribution of the AGN to the total IR luminosity is about one order of magnitude larger. A 5\% contribution to the total IR luminosity would imply that $\log (L^{\rm\,AGN}_{\rm\,IR}/\rm\,erg\,s^{-1})=43.66$ and the intrinsic 2--10\,keV luminosity of the AGN would be $\log (L_{2-10}/\rm\,erg\,s^{-1})=43.12$, also an order of magnitude larger than predicted by our X-ray spectral analysis. To have such a luminosity, the AGN should be obscured by $\log (N_{\rm\,H}/\rm cm^{-2})> 25$, which is inconsistent with the results obtained here. An alternative explanation for this discrepancy is that the AGN is intrinsically weak at X-ray wavelengths, as recently found by {\it NuSTAR} for the AGN in Mrk\,231 (\citealp{Teng:2014oq}, see also \citealp{Teng:2015vn}).

Assuming a 2--10\,keV bolometric correction of $\kappa_{\mathrm{x}}=20$ (e.g., \citealp{Vasudevan:2007fk}), the bolometric output of the AGN would be $7-40\times 10^{42}\rm\,erg\,s^{-1}$. This implies that the ratio between the IR luminosity and the total output of the AGN is $L^{\rm\,Bol}_{\rm\,AGN}/L_{\rm\,IR}\simeq 0.8-4.5\%$. The AGN bolometric output can also be inferred from the [Ne\,V]\,$14.32\mu$m luminosity, following the relation obtained by \cite{Satyapal:2007nx}:
\begin{equation}\label{eq:neV}
\log L_{\rm\,Bol}^{\rm\,AGN}=0.938\log L_{\rm\,[Ne\,V]}+6.317,
\end{equation}
and is $\log (L_{\rm\,Bol}^{\rm\,AGN}/\rm\,erg\,s^{-1})\sim 42.7$, consistent with the estimate obtained using the X-ray luminosity. The 2--10\,keV bolometric correction obtained using this value is $\kappa_{\mathrm{x}}\simeq 3-17$. 
The black hole mass of NGC\,6286 has been estimated to be $M_{\rm\,BH}\sim 2.7\times 10^{8}\,M_{\odot}$ by \cite{Caramete:2010zr} using the black hole mass-spheroid correlation (e.g., \citealp{Magorrian:1998ly}). The Eddington ratio of the source would then be $\lambda_{\rm\,Edd}\simeq (0.2-1.2)\times10^{-3}$, consistent with a low accretion rate AGN.

\begin{figure}[t!]
\centering
\includegraphics[width=8.5cm]{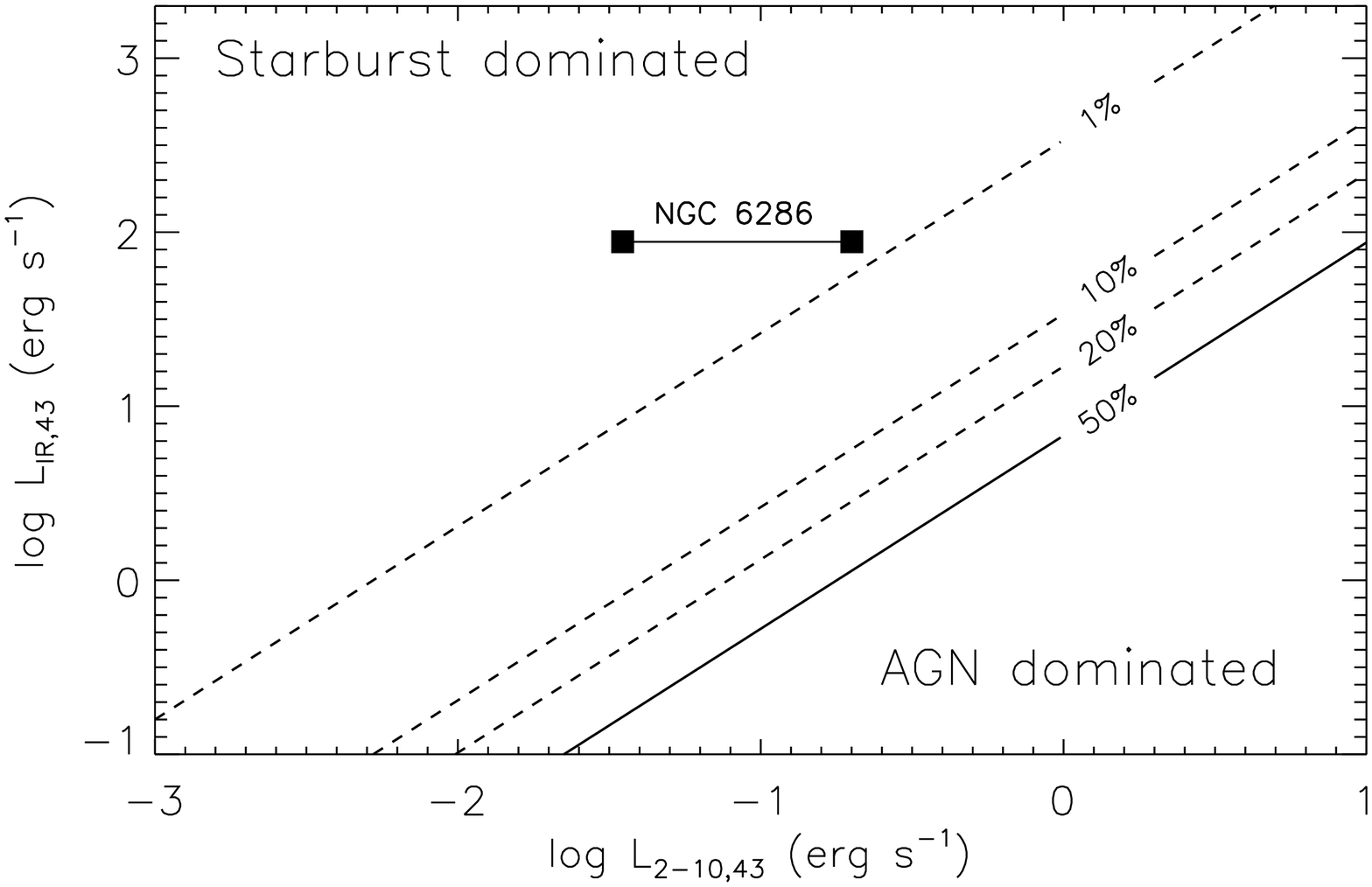}
  \caption{Intrinsic X-ray luminosity of the AGN in the 2--10\,keV band versus the total IR luminosity of the system (in the 8--1000$\mu$m band). Both luminosities are in units of $10^{43}\rm\,erg\,s^{-1}$. The continuous black line represents the values for which AGN and starburst contribute in the same amount to the IR flux, while the dashed lines show contributions of the AGN of 20, 10 and 1\%. The values of the IR luminosity expected to be due to the AGN are calculated from the 2--10\,keV luminosity following Eq.\,\ref{eq:xtoIR}. The two values of the 2--10\,keV luminosity of NGC\,6286 represent the minimum and maximum value obtained with the different models discussed in Sect\,\ref{sect:allxray} (see also Table\,\ref{tab:X-rayresults}). The figure shows that the AGN in NGC\,6286 contributes $<1\%$ of the total IR luminosity. }
\label{fig:lirtot_lx}
\end{figure}

The lack of a significant AGN contribution to the total IR flux is also confirmed considering the [Ne\,V]/[Ne\,II] ratio versus the EW of the 6.2$\mu$m PAH feature [see Fig.\,1 and 2 of \cite{Petric:2011zt}], which shows that the ratio between $L^{\rm\,AGN}_{\rm\,IR}$ and $L_{\rm\,IR}$ is below 1\% for this object. This, together with the 2--10\,keV bolometric correction obtained using [Ne\,V]\,$14.32\mu$m, clearly disfavours the intrinsically X-ray weak AGN scenario. We can therefore conclude that the energetics of NGC\,6286 are clearly dominated by the host galaxy, with the low-luminosity AGN providing only a minor contribution to the total flux. The contribution of the AGN to the IR flux of the system is shown in Fig.\,\ref{fig:lirtot_lx}.

\subsection{Optical and radio emission}\label{sect:optradio}
NGC\,6286 has been classified as a low-ionization nuclear emission-line region (LINER) by \cite{Veilleux:1987fk} using a classification scheme based on the diagram first proposed by Baldwin, Phillips \& Terlevich (\citeyear{Baldwin:1981kx}). While most LINERs appear to be driven by old stellar populations (e.g., \citealp{Sarzi:2010fk}) and by shocks in ULIRGs (e.g., \citealp{Soto:2010vn,Soto:2012uq}), in some cases they can be associated to low-luminosity AGN (e.g., \citealp{Ho:2008ly}). \cite{Yuan:2010ye} have recently used a new semi-empirical optical spectral classification to classify IR-selected galaxies based on three diagrams: [OIII]/H$\beta$ versus [NII]/H$\alpha$, [SII]/H$\alpha$ and [OI]/H$\alpha$ line ratios. This is based on the work of \cite{Kewley:2006uq} to separate starburst galaxies, starburst/AGN composite galaxies, Seyfert\,2s, and LINERs. In the scheme of \cite{Kewley:2006uq} objects that were classified as LINERs according to \cite{Veilleux:1987fk} would be either true LINERs, Seyfert\,2s, composite HII-AGN galaxies, or high metallicity star-forming galaxies. \cite{Yuan:2010ye} found that true LINERs are rare in IR-selected samples ($<5\%$), and most of the objects would be either classified as star-forming galaxies or starburst/AGN composites. \cite{Yuan:2010ye} classified NGC\,6286 as a composite using [NII], a HII region using [SII] and a LINER using [OI]. Therefore they adopted a composite classification for the source, which might imply the presence of an AGN. \cite{Yuan:2010ye} found that in the IR luminosity bin $L_{\rm\,IR}=10^{11}-10^{12}L_{\odot}$ about $37$\% of the objects in the {\it IRAS} Bright Galaxy Sample (BGS, \citealp{Sanders:1995kl}, \citealp{Veilleux:1995if}) are classified as composites.

To characterise the relative AGN contribution to the extreme ultraviolet (EUV) radiation field, \cite{Yuan:2010ye} use $D_{\rm\,AGN}$, which is the normalised distance from the outer boundary of the star-forming sequence.
While this quantity does not provide information on the fraction of emission due to the AGN, it can be used to compare the amount of EUV radiation due to the AGN in different objects.
For NGC\,6286 they found $D_{\rm\,AGN}=0.5$ using both the [OI]/$\rm H_{\alpha}$ and the [NII]/$\rm H_{\alpha}$ diagram. \cite{Yuan:2010ye} found a statistically significant increase of  $D_{\rm\,AGN}$ with $L_{\rm\,IR}$, consistent with the idea that the fraction of AGN increases for increasing values of the 8-100$\mu$m luminosity (e.g., \citealp{Veilleux:1995if}). The value obtained for NGC\,6286 is marginally larger than the average value obtained by \cite{Yuan:2010ye} for the BGS sample for $11<\log (L_{\rm\,IR}/L_{\odot})<12$ ($D_{\rm\,AGN}\simeq 0.35$).

The dense molecular gas tracer HCN has been found to be enhanced (relative to HCO+ and CO) in systems with dominant AGN (e.g., \citealp{Imanishi:2007fv}). \citet{Privon:2015dz} have shown that some pure starburst and composite sources show similarly enhanced HCN emission. The origin of this enhancement is uncertain, but might be due to mid-infrared pumping associated with a compact obscured nucleus (CON; e.g., \citealp{Aalto:2015fu}). However, the HCN/HCO+ ratio of NGC\,6286 is consistent with that of normal starbursts, rather than CONs. From this we can conclude that the starburst does not appear to be compact.

A radio core is rather common in low-luminosity AGN, as shown by the work of \cite{Nagar:2005zr}, who found evidence of radio emission in $\geq 50\%$ of the low-luminosity AGN of the Palomar Spectroscopic sample (see also \citealp{Ho:2008ly}). The flux of NGC\,6286 at 1.4\,GHz is $f_{1.4\rm\,GHz}=157.4\pm5.6$\,mJy \citep{Condon:1998ij}, which implies that the radio loudness is $\log R_{\rm\,X}=\log (f_{1.4\rm\,GHz}/f_{2-10})=-2.6$ to $-3.1$, depending on the X-ray spectral model assumed. These values were obtained taking into account only the nuclear emission in the computation of the 2--10\,keV flux. Considering the threshold suggested by \citet{La-Franca:2010uq} (see also \citealp{Panessa:2007qf}, \citealp{Terashima:2003bh}), $\log R_{\rm\,X}=-4.3$, NGC\,6286 would be classified as radio-loud AGN. 
\cite{Murphy:2013ve} report that the radio spectral index\footnote{We consider here the following definition of the spectral index: $F_{\nu}\propto \nu^{\alpha}$.} of NGC\,6286 is $\alpha_{\rm\,low}=-0.73\pm0.03$, $\alpha_{\rm\,mid}=-0.89\pm0.03$ and $\alpha_{\rm\,high}=-1.02\pm0.12$ for $\nu <5$\,GHz, $1<\nu/\rm GHz<10$ and $\nu>10$\,GHz, respectively. This would point towards a significant contribution of synchrotron emission, possibly from a jet. The two radio sources detected by EVN and coincident with the 3--8\,keV core could be in fact associated to a jet and counter jet, consistent with the radio-loud classification of NGC\,6286.

\subsection{Heavily obscured low-luminosity AGN in U/LIRGs}\label{sect:obscuredAGNinLIRGs}
As discussed above for the case of NGC\,6286, the identification of heavily obscured AGN in LIRGs can be rather difficult if the AGN has a low-luminosity. The EW of PAH features would not be significantly affected by the AGN if it is highly obscured, since the gas and dust would shield the PAH-emitting molecules, or if it is not very luminous. A low-luminosity AGN would also be difficult to find by studying the 2.5--5$\mu$m slope, since the IR emission would be dominated by the starburst, and the AGN emission can still be self-absorbed. Absorption features also would not be able to help if the AGN is not very luminous. A more reliable tracer is [Ne\,V], but while its detection might indicate the presence of an AGN, its non-detection does not exclude it. Moreover, [Ne\,V] could be created in young starbursts, and for low-luminosity AGN it could be too faint to be detected (see Eq.\,\ref{eq:neV}). Radio studies can also give important insights, but since not all AGN are very strong at these wavelengths, results are not always conclusive. Hard X-ray studies are possibly the best way to unveil obscured AGN in U/LIRGs, although they can also be limited by absorption for $\log (N_{\rm\,H}/\rm cm^{-2})\gg 24$.

By using multi-wavelength indicators of AGN for a subsample of 53 U/LIRGs within the GOALS sample, \cite{U:2012fc} found that $\sim 60\%$ and $\sim 25\%$ of ULIRGs and LIRGs host AGN. Studying the whole GOALS sample, \cite{Petric:2011zt} found that 18\% of the LIRGs show evidence of [Ne\,V]\,$14.32\mu$m, and hence might host an AGN. By means of optical spectroscopy, \cite{Yuan:2010ye} found that $59\%$ of the 51 single nuclei galaxies with $11<\log (L_{\rm\,IR}/L_{\odot})<12$ in the BGS sample host an AGN\footnote{37\% are composite AGN/starburst, 14\% Seyfert\,2s, 2\% Sy1s and 6\% LINERs.}. The fraction of AGN is larger (77\%) if one considers only two of the three diagrams for the spectral classification. A significant fraction of the composite systems might hide buried low-luminosity AGN, as in the case of NGC\,6286, although an important contribution to the line emission in these objects might be due to shocks (e.g., \citealp{Soto:2012kx}). \cite{Treister:2010zr} have shown, by stacking {\it Chandra} spectra of LIRGs in the {\it Chandra} Deep Field-South, that 15\% of the objects with $L_{\rm\,IR}>10^{11}\,L_{\odot}$ contain heavily obscured AGN. By stacking X-ray spectra in different bins of stellar mass, they found a significant excess at $E=6-7$\,keV in the stacked spectrum of sources with mass $M>10^{11}M_{\odot}$, very likely related to a prominent Fe K$\alpha$ line, while no clear evidence of AGN activity was found in less-massive galaxies. \cite{Treister:2010zr} concluded that there might be a large population of heavily obscured AGN in high mass galaxies. NGC\,6286, with a stellar mass of $1.26\times10^{11}M_{\odot}$ \citep{Howell:2010uq}, fits extremely well into this scenario in the local Universe.

We have shown in $\S$\,\ref{sect:tracers} and \ref{sect:optradio} that NGC\,6286 has optical and IR characteristics quite typical of LIRGs, and consistent with other galaxies of the GOALS sample for the same merger stage. It is interesting to notice that also the hardness ratio and the observed 2--10\,keV luminosity inferred by {\it Chandra} are consistent with a large fraction of the objects of the sample of \cite{Iwasawa:2011fk} (see Fig.\,5 and 6 of their paper, respectively), which might indicate that several more heavily obscured low-luminosity AGN are present in LIRGs of the GOALS sample. Moreover, we have shown that in the low-count regime it is possible to miss obscured AGN by adopting a simple phenomenological model to reproduce their X-ray spectra. Therefore there might be a significant population of low-luminosity heavily obscured AGN in LIRGs that we are missing due to the lack of sensitive hard X-ray observations. Numerical simulations have shown that accretion onto SMBHs might be happen at some level even after the first encounter (e.g., \citealp{Di-Matteo:2005qf}), although the expected accretion rate varies depending on the galaxy mergers code adopted (e.g., \citealp{Gabor:2015ve}). Our on-going campaign of {\it NuSTAR} observations of ten LIRGs will allow us to study the AGN fraction in merging galaxies in the hard X-ray band across the whole merger sequence. 

Another object showing similar characteristics to NGC\,6286 is IC\,883, a LIRG in a late merger stage that was found to host a low-luminosity AGN from radio observations (\citealp{Romero-Canizales:2012fk}; Romero-Ca$\mathrm{\tilde{n}}$izales et al., in prep.). As for NGC\,6286, the IR emission of IC\,883 is dominated by star-formation and the AGN does not contribute significantly to the energetics of the system. Interestingly, similar to NGC\,6268, IC\,883 is also reported as a composite AGN/starburst system by \cite{Yuan:2010ye}, along with more than one third of LIRGs from the BGS sample.

\section{Summary and conclusions}\label{section:summary}
We have reported here the first results of a {\it NuSTAR} campaign aimed at observing ten LIRGs in different merger stages, focussing on the first detection of a heavily obscured AGN in NGC\,6286. The {\it Chandra}/ACIS-S 0.3--2\,keV image of the source shows extended emission that covers $\sim 4.4$\,kpc (Fig.\,\ref{fig:images}), and which might be due to collisionally ionized plasma. In the 3--8\,keV band we found a compact source, with a flat 1.2--8\,keV spectrum ($\Gamma \sim -0.2$), which coincides with the radio emission detected by FIRST. The {\it NuSTAR} spectrum also shows a flat X-ray continuum ($\Gamma\sim 0.5$). By analysing the broad-band X-ray spectrum of the source, combining archival {\it XMM-Newton}, {\it Chandra} and quasi-simultaneous {\it NuSTAR} and {\it Swift}/XRT observations, we have found that the source is consistent with being obscured by mildly CT material ($N_{\rm\,H}=1.08^{+0.63}_{-0.38}\times 10^{24}\rm\,cm^{-2}$, Fig.\,\ref{fig:dchiNH}). The presence of a heavily obscured AGN is confirmed by the possible detection of weak [Ne\,V]\,$14.32\mu$m and [Ne\,V]\,$24.32\mu$m lines \citep{Dudik:2009cq}, by near-IR to radio spectral decomposition \citep{Vega:2008kh} and by the optical classification of the galaxy as an AGN/starburst composite \citep{Yuan:2010ye}.

The buried AGN has an intrinsically low luminosity ($L_{2-10}\sim 3-20\times 10^{41}\rm\,erg\,s^{-1}$), a low value of the Eddington ratio [$\lambda_{\rm\,Edd}\simeq (0.2-1.2)\times10^{-3}$] and seems to contribute less than 1\% to the energetics of the system (Fig.\,\ref{fig:lirtot_lx}). Because of its low luminosity, previous observations carried out below 10\,keV and in the infrared did not notice the presence of a buried AGN. By exploiting the rich multi-wavelength coverage of U/LIRGs in the GOALS sample we have discussed the radio to X-ray characteristics of NGC\,6286, showing that they are consistent with those of objects with similar luminosities and in a similar merger stage. We speculated that there might be a significant fraction of low-luminosity AGN in LIRGs that we are missing due to their low contribution to the total IR flux of the system. In particular, more than one third of LIRGs from the BGS sample are classified as composite AGN/starburst system by \cite{Yuan:2010ye}, and might hide low-luminosity highly obscured AGN.

The case of NGC\,6286 clearly shows the importance of hard X-ray coverage in order to detect low-luminosity heavily obscured AGN in LIRGs. Our ongoing {\it NuSTAR} campaign of LIRGs will put better constraints on the fraction of these objects and the relation between obscured accretion and merger stage.

\begin{table*}
\begin{center}
\caption[]{Summary of the X-ray spectral analysis for the spatially integrated X-ray spectrum of NGC\,6286.}
\label{tab:X-rayresults}
\begin{tabular}{lccc}
\noalign{\smallskip}
\hline \hline \noalign{\smallskip}
\noalign{\smallskip}
\multicolumn{4}{c}{\textbf{PEXRAV}}   \\
\noalign{\smallskip}
\hline \noalign{\smallskip}
	 	Column density ($N_{\rm\,H})$	[\scriptsize{$10^{22}\rm\,cm^{-2}$}]									&    \multicolumn{3}{c}{	$132^{+82}_{-54}$}	\\
\noalign{\smallskip}
	 	Reflection parameter ($R$)										&    \multicolumn{3}{c}{	$\leq 0.4$}	\\
\noalign{\smallskip}
 	Plasma temperature ($kT$)	[\scriptsize{$\rm\,keV$}]								&    \multicolumn{3}{c}{	$0.79^{\,A}$}	\\
\noalign{\smallskip}
	  	Scattered fraction ($f_{\rm\,scatt}$)		[\scriptsize{$\%$}]						&    \multicolumn{3}{c}{	$1.3^{+1.9}_{-0.7}$$^{\,B}$}	\\
\noalign{\smallskip}
 	Fe K$\alpha$ EW		[\scriptsize{$\rm\,eV$}]										&    \multicolumn{3}{c}{	$\leq 2318$}	\\
\noalign{\smallskip}
 	Observed 2--10\,keV flux ($F_{2-10}^{\rm\,obs}$)	[\scriptsize{$10^{-13}\rm\,erg \,s^{-1}\,cm^{-2}$}]								&   \multicolumn{3}{c}{ 	$0.9$}	\\
\noalign{\smallskip}
	 	Observed 10--50\,keV flux ($F_{10-50}^{\rm\,obs}$)		[\scriptsize{$10^{-13}\rm\,erg \,s^{-1}\,cm^{-2}$}]							&   \multicolumn{3}{c}{ 	$9.6$}	\\
\noalign{\smallskip}
	 	Intrinsic 2--10\,keV flux ($F_{2-10}$)		[\scriptsize{$10^{-12}\rm\,erg \,s^{-1}\,cm^{-2}$}]							&    \multicolumn{3}{c}{	$2.6$}	\\
\noalign{\smallskip}
	 	intrinsic 10--50\,keV flux ($F_{10-50}$)	[\scriptsize{$10^{-12}\rm\,erg \,s^{-1}\,cm^{-2}$}]								&    \multicolumn{3}{c}{	$3.1$}	\\
\noalign{\smallskip}
	  	2--10\,keV luminosity ($L_{2-10}$)		[\scriptsize{$\rm\,erg \,s^{-1}$}]							&   \multicolumn{3}{c}{ 	$2.01\times 10^{42}$}	\\
\noalign{\smallskip}
	  	10--50\,keV luminosity ($L_{10-50}$)		[\scriptsize{$\rm\,erg \,s^{-1}$}]								&    \multicolumn{3}{c}{	$2.34\times10^{42}$}	\\
\noalign{\smallskip}
	  	$\chi^2$/DOF								&    \multicolumn{3}{c}{	47.2/44}	\\
\noalign{\smallskip}
\hline
\noalign{\smallskip}
%
%
%
%
\noalign{\smallskip}
\multicolumn{4}{c}{\textbf{TORUS}}   \\
\noalign{\smallskip}
\hline
\noalign{\smallskip}
\multicolumn{1}{c}{ } & \multicolumn{1}{c}{$\theta_{\rm\,OA}=40^{\circ}$}  & \multicolumn{1}{c}{$\theta_{\rm\,OA}=60^{\circ}$} & \multicolumn{1}{c}{$\theta_{\rm\,OA}=80^{\circ}$}  \\
\noalign{\smallskip}
\hline
\noalign{\smallskip}
 	Plasma temperature ($kT$)	[\scriptsize{$\rm\,keV$}]												&    \multicolumn{1}{c}{	$0.79^{A}$} 		&    \multicolumn{1}{c}{	$ 0.79^{A}$} 				&    \multicolumn{1}{c}{	$ 0.79^{A}$}	\\
\noalign{\smallskip}
	  	Scattered fraction ($f_{\rm\,scatt}$)		[\scriptsize{$\%$}]										&    \multicolumn{1}{c}{$3.0^{+2.0}_{-1.3}$$^B$} &    \multicolumn{1}{c}{	$2.4 ^{+2.0}_{-1.3}$$^B$} 	&    \multicolumn{1}{c}{	$ 2.0^{+1.9}_{-1.3}$$^B$}	\\
\noalign{\smallskip}
	 		Column density ($N_{\rm\,H})$	[\scriptsize{$10^{22}\rm\,cm^{-2}$}]								&    \multicolumn{1}{c}{	$110^{+89}_{-39}$} &    \multicolumn{1}{c}{	$ 111^{+109}_{-41}$} 			&    \multicolumn{1}{c}{	$ 106^{+101}_{-38}$}	\\
\noalign{\smallskip}
	 	Intrinsic 2--10\,keV flux ($F_{2-10}$)		[\scriptsize{$10^{-13}\rm\,erg \,s^{-1}\,cm^{-2}$}]				&    \multicolumn{1}{c}{	$ 10.7$} 			&    \multicolumn{1}{c}{	$13.0$} 					&    \multicolumn{1}{c}{	$16.0$}	\\
\noalign{\smallskip}
	 	intrinsic 10--50\,keV flux ($F_{10-50}$)	[\scriptsize{$10^{-13}\rm\,erg \,s^{-1}\,cm^{-2}$}]				&    \multicolumn{1}{c}{	$13.1$} 			&    \multicolumn{1}{c}{	$15.8$} 					&    \multicolumn{1}{c}{	$19.5$}	\\
\noalign{\smallskip}
	  	2--10\,keV luminosity ($L_{2-10}$)		[\scriptsize{$\rm\,erg \,s^{-1}$}]								&   \multicolumn{1}{c}{ 	$8.14 \times 10^{41}$} &   \multicolumn{1}{c}{ 	$9.84 \times 10^{41}$} 		&   \multicolumn{1}{c}{ 	$ 1.21\times 10^{42}$}	\\
\noalign{\smallskip}
	  	10--50\,keV luminosity ($L_{10-50}$)		[\scriptsize{$\rm\,erg \,s^{-1}$}]								&    \multicolumn{1}{c}{	$ 9.90\times10^{41}$}&   \multicolumn{1}{c}{ 	$1.20\times 10^{42}$} 		&   \multicolumn{1}{c}{ 	$ 1.48\times 10^{42}$}	\\

\noalign{\smallskip}
	$\chi^2$/DOF								&    \multicolumn{1}{c}{47.8/46} &    \multicolumn{1}{c}{	48.1/46}  &    \multicolumn{1}{c}{48.5/46}	\\
\noalign{\smallskip}
\hline
\noalign{\smallskip}
\noalign{\smallskip}
\multicolumn{4}{c}{\textbf{SPHERE}}   \\
\noalign{\smallskip}
\hline
\noalign{\smallskip}
 	Plasma temperature ($kT$)	[\scriptsize{$\rm\,keV$}]								&    \multicolumn{3}{c}{	$0.79^{\,A}$}	\\
\noalign{\smallskip}
	  	Scattered fraction ($f_{\rm\,scatt}$)		[\scriptsize{$\%$}]						&    \multicolumn{3}{c}{	$3.6^{+2.1}_{-1.2}$$^B$}	\\
\noalign{\smallskip}
	 		Column density ($N_{\rm\,H})$	[\scriptsize{$10^{22}\rm\,cm^{-2}$}]									&       \multicolumn{3}{c}{	$101^{+46}_{-32}$}	\\
\noalign{\smallskip}
	 	Intrinsic 2--10\,keV flux ($F_{2-10}$)		[\scriptsize{$10^{-13}\rm\,erg \,s^{-1}\,cm^{-2}$}]							&    \multicolumn{3}{c}{	$9.2$}	\\
\noalign{\smallskip}
	 	intrinsic 10--50\,keV flux ($F_{10-50}$)	[\scriptsize{$10^{-13}\rm\,erg \,s^{-1}\,cm^{-2}$}]								&    \multicolumn{3}{c}{	$11.2$}	\\
\noalign{\smallskip}
	  	2--10\,keV luminosity ($L_{2-10}$)		[\scriptsize{$\rm\,erg \,s^{-1}$}]							&   \multicolumn{3}{c}{ 	$6.98\times10^{41}$}	\\
\noalign{\smallskip}
	  	10--50\,keV luminosity ($L_{10-50}$)		[\scriptsize{$\rm\,erg \,s^{-1}$}]								&    \multicolumn{3}{c}{	$8.50\times10^{41}$}	\\
\noalign{\smallskip}
	$\chi^2$/DOF								&    \multicolumn{3}{c}{	47.2/46} 	\\
\noalign{\smallskip}
\hline
\noalign{\smallskip}
\noalign{\smallskip}
\multicolumn{4}{c}{\textbf{MYTORUS}}   \\
\noalign{\smallskip}
\hline
\noalign{\smallskip}
 	Photon index ($\Gamma$)									&    \multicolumn{3}{c}{	$1.53^{+0.10}_{-NC}$$^C$}	\\
\noalign{\smallskip}
 	Plasma temperature ($kT$)	[\scriptsize{$\rm\,keV$}]								&    \multicolumn{3}{c}{	$0.79^{\,A}$}	\\
\noalign{\smallskip}
	  	Scattered fraction ($f_{\rm\,scatt}$)		[\scriptsize{$\%$}]						&    \multicolumn{3}{c}{	$ 8.8^{+4.9}_{-2.6}$$^B$}	\\
\noalign{\smallskip}
	 		Column density ($N_{\rm\,H})$	[\scriptsize{$10^{22}\rm\,cm^{-2}$}]									&       \multicolumn{3}{c}{	$95^{+61}_{-39}$}	\\
\noalign{\smallskip}
	 	Intrinsic 2--10\,keV flux ($F_{2-10}$)		[\scriptsize{$10^{-13}\rm\,erg \,s^{-1}\,cm^{-2}$}]							&    \multicolumn{3}{c}{	$4.7$}	\\
\noalign{\smallskip}
	 	intrinsic 10--50\,keV flux ($F_{10-50}$)	[\scriptsize{$10^{-13}\rm\,erg \,s^{-1}\,cm^{-2}$}]								&    \multicolumn{3}{c}{	$11.1$}	\\
\noalign{\smallskip}
	  	2--10\,keV luminosity ($L_{2-10}$)		[\scriptsize{$\rm\,erg \,s^{-1}$}]							&   \multicolumn{3}{c}{ 	$3.49\times10^{41}$}	\\
\noalign{\smallskip}
	  	10--50\,keV luminosity ($L_{10-50}$)		[\scriptsize{$\rm\,erg \,s^{-1}$}]								&    \multicolumn{3}{c}{	$8.35\times10^{41}$}	\\
\noalign{\smallskip}
	$\chi^2$/DOF								&    \multicolumn{3}{c}{	52.5/45} 	\\
\noalign{\smallskip}
\hline
\noalign{\smallskip}
\multicolumn{4}{l}{{\bf Notes}. $^A$ parameter left free to vary within the uncertainties of the value obtained fitting}\\
\multicolumn{4}{l}{the extended emission (see $\S$\,\ref{sect:allxray} for details). $^B$ value calculated from the ratio of $n_{\rm\,po}$ and $n_{\rm\,po}^{\rm\,scatt}$.}\\
\multicolumn{4}{l}{$^C$ the photon index in \textsc{MyTORUS} is calculated only in the range $\Gamma=1.4-2.6$.}

\end{tabular}
\end{center}
\end{table*}

\acknowledgments
We thank the anonymous referee for his/her comments, that helped us to improve the quality of our manuscript, and the {\it NuSTAR} Cycle 1 TAC for the {\it NuSTAR} data on which this paper is based. CR acknowledges C.S. Chang, H. Inami, P. Gandhi and S. Satyapal for useful discussion. We thank Adam Block (Mount Lemmon SkyCenter/University of Arizona) for allowing us to publish his optical image of NGC\,6286/NGC\,6285.
This research has made use of the {\it NuSTAR} Data Analysis Software (\textsc{NuSTARDAS}) jointly developed by the ASI Science Data Center (ASDC, Italy) and the California Institute of Technology (Caltech, USA), and of the NASA/ IPAC Infrared Science Archive and NASA/IPAC Extragalactic Database (NED), which are operated by the Jet Propulsion Laboratory, California Institute of Technology, under contract with the National Aeronautics and Space Administration.
We acknowledge financial support from the CONICYT-Chile grants ``EMBIGGEN" Anillo ACT1101 (CR, FEB, ET), FONDECYT 1141218 (CR, FEB),  FONDECYT 315238 (CRC), FONDECYT 3150361 (GP), Basal-CATA PFB--06/2007 (CR, FEB, ET), the NASA {\it NuSTAR} AO1 Award NNX15AV27G (FEB) and the Ministry of Economy, Development, and Tourism's Millennium Science Initiative through grant IC120009, awarded to The Millennium Institute of Astrophysics, MAS (FEB, CRC). KS gratefully acknowledges support from Swiss National Science Foundation Grant PP00P2\_138979/1. MI was supported by JSPS KAKENHI Grant Number 23540273 and 15K05030.

{\it Facilities:} \facility{NuSTAR}, \facility{Chandra}, \facility{Swift}, \facility{XMM-Newton}.


\bibliographystyle{apj} 
\bibliography{NGC6286_ref.bib}

\end{document}